\documentclass[10pt,aps,prx,twocolumn,notitlepage,showpacs,superscriptaddress,longbibliography]{revtex4-2}

\usepackage{times}
\usepackage{amssymb,amsmath}
\usepackage{bm}
\usepackage{graphicx}
\usepackage{graphicx,color}

\usepackage[urlcolor=blue,colorlinks=true,citecolor=blue,linkcolor=blue,pdfstartview={FitH},bookmarks=false]{hyperref}
\begin{document}
	
\title{Microscopic Mechanism of Pair-, Charge- and Spin-Density-Wave Instabilities in Interacting \boldmath{$D$}-Dimensional Fermi Liquids}

\date{\today}
\author{Dmitry Miserev,$^{1\ast}$ Herbert Schoeller,$^2$ Jelena Klinovaja,$^{1}$ and Daniel Loss}
\affiliation{Department of Physics, University of Basel, \\
	Klingelbergstrasse 82, CH-4056 Basel, Switzerland\\
	$^2$Institut f\"{u}r Theorie der Statistischen Physik, RWTH Aachen University and JARA -- Fundamentals of Future Information Technology,\\
	52056 Aachen, Germany}

\begin{abstract}
	We present an analytic theory unraveling the microscopic mechanism of instabilities within interacting $D$-dimensional Fermi liquid. 
	Our model consists of a $D$-dimensional electron gas subject to an instantaneous electron-electron interaction of a finite range exceeding the average inter-particle distance.  
	Pair, charge and spin susceptibilities are evaluated via the one-loop renormalization group theory and via the bosonization approach, giving identical results.
	In case of a repulsive interaction, 
	we identify an intrinsic Fermi liquid instability towards insulating spin/charge density wave order when the interaction coupling strength reaches a universal critical value. 
	If both electron and hole pockets of the same size are present, the ground state is an excitonic insulator at arbitrarily small repulsive interaction.
	If the interaction is attractive, the ground state is a singlet non-BCS superconductor with a uniform condensate.
	In case if both electron and hole Fermi surfaces are present, we predict an instability towards the inter-pocket pair-density-wave ordering at the critical coupling.
	This prediction lends strong theoretical support to the pair-density-wave scenario of superconductivity in cuprate materials.
	Due to its simple and universal nature, presented microscopic mechanism of intrinsic instabilities of interacting $D$-dimensional Fermi liquids constitutes a solid theoretical ground for understanding quantum phase transitions in a variety of quantum materials, from ultraclean semiconductor quantum wells to high-temperature superconductors.
\end{abstract}

\maketitle	

\section{Introduction}
Strongly correlated phenomena observed in various condensed matter systems have long posed a perplexing challenge to the condensed matter physics community \cite{agterbergPhysicsPairDensityWaves2020,satoObservationDx2y2LikeSuperconducting2001,doiron-leyraudQuantumOscillationsFermi2007,helmEvolutionFermiSurface2009,jangSuperconductivityInsensitiveOrderEnsuremath2017,liHolePocketDriven2019,fournierInsulatorMetalCrossoverOptimal1998,huckerCompetingChargeSpin2014,satoThermodynamicEvidenceNematic2017,mokashiCriticalBehaviorStrongly2012,hossainObservationSpontaneousFerromagnetism2020,melnikovQuantumPhaseTransition2019,shashkinMetalInsulatorTransition2021,caoCorrelatedInsulatorBehaviour2018,sharpeEmergentFerromagnetismThreequarters2019,xieSpectroscopicSignaturesManybody2019,wongCascadeElectronicTransitions2020,jiangChargeOrderBroken2019,luSuperconductorsOrbitalMagnets2019,yankowitzTuningSuperconductivityTwisted2019,caoUnconventionalSuperconductivityMagicangle2018,ohEvidenceUnconventionalSuperconductivity2021,parkTunableStronglyCoupled2021,aroraSuperconductivityMetallicTwisted2020,wangCorrelatedElectronicPhases2020,shenCorrelatedStatesTwisted2020,rickhausCorrelatedElectronholeState2021,suSuperconductivityTwistedDouble2023}. 
While existing theoretical models have contributed valuable insights into individual phases \cite{kohnNewMechanismSuperconductivity1965,baranovSuperconductivitySuperfluidityFermi1992,chubukovMagnetismSuperconductivityPairing2008,raghuSuperconductivityRepulsiveInteractions2011,chubukovPairingMechanismFeBased2012,chubukovMagnetismSuperconductivitySpontaneous2016,khveshchenkoLowenergyPropertiesTwodimensional1993,leeAmpereanPairingInstability2007,leeAmpereanPairingPseudogap2014,isobeUnconventionalSuperconductivityDensity2018,laksonoSingletSuperconductivityEnhanced2018,liuChiralSpinDensity2018,sachdevGaplessSpinfluidGround1993,altshulerLowenergyPropertiesFermions1994,wangSuperconductivityQuantumCriticalPoint2016,chowdhuryTranslationallyInvariantNonFermiLiquid2018,abanovInterplaySuperconductivityNonFermi2020,panYukawaSYKModelSelftuned2021,chowdhurySachdevYeKitaevModelsWindow2022,davisonHolographicDualityResistivity2014,andradeCoherentVsIncoherent2019,andradePhaseRelaxationPattern2021,balmTlinearResistivityOptical2023}, a consensus on the microscopic mechanism underlying strongly correlated electron matter within a realistic model is lacking \cite{agterbergPhysicsPairDensityWaves2020}.

\begin{figure}[t]
	\centering
	\includegraphics[width=0.95\columnwidth]{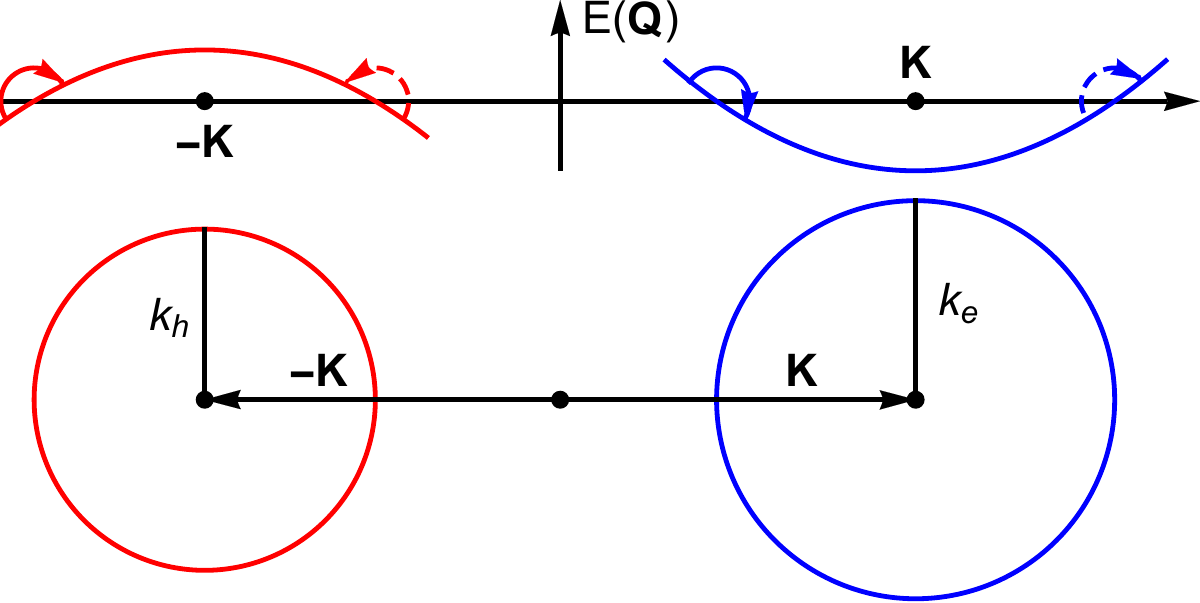}
	\caption{{\bf Particle spectrum and Fermi surfaces.}
	$D$-dimensional, spin degenerate, electron (blue) and hole (red) Fermi pockets with the Fermi momenta $k_e$ and $k_h$ centered at $\bm K$ and $-\bm K$, respectively. 
	An alternative choice of the Brillouin zone origin does not impact any physical observables. 
	The energy vs. momentum $\bm Q$, $E(\bm Q)$, is a sketch of the particle dispersion along the line connecting the centers of the two pockets. 
	Blue and red solid arrows indicate the inter-pocket particle-hole scattering process due to repulsive electron-electron interaction of finite range $R_0$ responsible for the excitonic insulator phase transition at momentum $2 \bm K + \bm q$, $|\bm q| = |k_e - k_h|$.
	Blue and red dashed arrows illustrate the inter-pocket pair scattering process due to attractive interactions leading to the PDW singularity at finite momentum $k_e + k_h$.}
	\label{fig:1}
\end{figure}

A phenomenological framework for investigating strongly correlated electron systems facilitates the exploration of various orders, encompassing charge-density-wave (CDW), spin-density-wave (SDW), and pair-density-wave (PDW) orders \cite{agterbergPhysicsPairDensityWaves2020,bergCharge4eSuperconductivityPairdensitywave2009,bergStripedSuperconductorsHow2009,agterbergCheckerboardOrderVortex2015,wangPairDensityWaves2018}.
To the best of our knowledge, a reliable microscopic mechanism explaining the origins of insulating and unconventional superconducting orders in $D$-dimensional quantum materials with $D > 1$ is still unclear \cite{agterbergPhysicsPairDensityWaves2020}.
In this paper, we aim to fill this gap and present a general microscopic model that demonstrates the instability of an interacting electron gas towards insulating CDW and SDW orders in case of a repulsive interaction and non-BCS superconducting order for an attractive interaction.
If both electron and hole pockets are present, see Fig.~\ref{fig:1}, we predict the PDW instability when attractive interaction reaches a universal critical value. 
This lends strong theoretical support for the PDW scenario of high-temperature superconductivity \cite{agterbergPhysicsPairDensityWaves2020,bergCharge4eSuperconductivityPairdensitywave2009,bergStripedSuperconductorsHow2009,agterbergCheckerboardOrderVortex2015,wangPairDensityWaves2018} observed in cuprate materials \cite{satoObservationDx2y2LikeSuperconducting2001,doiron-leyraudQuantumOscillationsFermi2007,helmEvolutionFermiSurface2009,jangSuperconductivityInsensitiveOrderEnsuremath2017,liHolePocketDriven2019,fournierInsulatorMetalCrossoverOptimal1998,huckerCompetingChargeSpin2014,satoThermodynamicEvidenceNematic2017}.

Importantly, the electron-electron interaction, $U(\tau, \bm r)$, within our model is instantaneous and assumed to have a finite range $R_0$ that significantly exceeds the average inter-electron distance given by the Fermi wavelength, $\lambda_F$,
\begin{eqnarray}
	&& U(\tau, \bm r) = \delta (\tau) V(r), \hspace{5pt} \lambda_F \ll r \lesssim R_0 , \label{inter}
\end{eqnarray}
where $\delta (\tau)$ is delta function.
In the repulsive case, $V(r) > 0$, Eq.~(\ref{inter}) models a screened Coulomb interaction, where $R_0$ plays role of the Thomas-Fermi screening length scaling as $\lambda_F \gamma^{-1/(D - 1)} \gg \lambda_F$ at weak interaction coupling $\gamma \ll 1$.
In the attractive case, $V(r) < 0$, such interaction models a propagator of fast bosons with typical frequency $\omega_0 \gtrsim v_F/R_0$, where $v_F$ is the Fermi velocity.
This limit is opposite to the BCS limit $\omega_0 \ll v_F/R_0$ where the Migdal theorem \cite{migdal1958} guarantees small corrections to the interaction vertex.
In contrast, in the non-BCS limit, $\omega_0 \gtrsim v_F/R_0$, considered here, the vertex corrections are strongly non-local and provide the leading contribution.
This strongly non-BCS limit is relevant to superconductivity induced by optical phonons, plasmons \cite{takadaPlasmonMechanismSuperconductivity1978,sharmaSuperconductivityCollectiveExcitations2020,grankinInterplayHyperbolicPlasmons2023}, spinons \cite{abanovQuantumcriticalTheorySpinfermion2003} and also by photons in cavity-coupled electron systems \cite{schlawinCavityMediatedElectronPhotonSuperconductivity2019}.

As it has been recently shown in Ref.~\cite{hutchinsonSpinSusceptibilityInteracting2023}, interactions in Eq.~(\ref{inter}) generate large logarithmic corrections to charge and spin susceptibilities already in first-order perturbation theory.
In contrast, a short-range (contact) interaction only generates irrelevant non-analyticities \cite{chubukovSingularPerturbationTheory2005,gangadharaiahInteractingFermionsTwo2005,aleinerSupersymmetricLowenergyTheory2006,maslovNonanalyticParamagneticResponse2009,zakSpinSusceptibilityInteracting2010,zakFerromagneticOrderNuclear2012}.
Here, we present pair, charge, and spin susceptibilities calculated within two different non-perturbative approaches.
The first approach is based on the one-loop renormalization group (RG) procedure \cite{menyhardApplicationRenormalizationGroup1973,solyomApplicationRenormalizationGroup1973} that allows us to calculate critical exponents at weak interaction coupling.
The second approach is based on multidimensional bosonization \cite{haldaneLuttingerTheoremBosonization2005,castronetoBosonizationFermiLiquids1994,frohlichEffectiveGaugeField1995,metznerFermiSystemsStrong1998,kopietzBosonizationInteractingFermions2006,efetovExactBosonizationInteracting2009,delacretazNonlinearBosonizationFermi2022} which is asymptotically exact in the semiclassical/infrared limit \cite{miserevDimensionalReductionLuttingerWard2023a}.
Both approaches agree in the weak coupling regime.
All calculations are performed on the Fermi liquid (FL) side at zero temperature.

In case of a repulsive finite-range interaction, singularities at the critical coupling are identified exclusively in the $2k_F$ harmonics of charge and spin susceptibilities, where $k_F$ is the Fermi momentum.
The pair susceptibility remains non-singular.
This behavior signals a phase transition from the FL to a correlated insulator state characterized by competing CDW and SDW orders. 
These findings are consistent with metal-to-insulator transitions observed experimentally in ultraclean semiconductor quantum wells at very low electron densities \cite{mokashiCriticalBehaviorStrongly2012,hossainObservationSpontaneousFerromagnetism2020,melnikovQuantumPhaseTransition2019,shashkinMetalInsulatorTransition2021}.
In case of multiple Fermi surfaces, also the inter-pocket charge and spin susceptibilities exhibit singularities at the critical coupling.
In particular, if electron and hole pockets are present, the ground state is an excitonic insulator \cite{jeromeExcitonicInsulator1967,khveshchenkoGhostExcitonicInsulator2001,herbutInteractionsPhaseTransitions2006,drutGrapheneVacuumInsulator2009}, see Fig.~\ref{fig:1}.
Excitonic insulators have been recently reported for a number of layered materials \cite{luZerogapSemiconductorExcitonic2017,maStronglyCorrelatedExcitonic2021,guDipolarExcitonicInsulator2022,jiaEvidenceMonolayerExcitonic2022}.

If the interaction is attractive, we find that the zero-momentum component of the pair susceptibility exhibits a non-BCS singularity at any value of the interaction coupling constant.
The non-BCS character of this superconductivity signature is, again, due to the instantaneous but finite-range interaction that models an electron-electron coupling via a fast boson with typical frequency $\omega_0 \gtrsim v_F/R_0$.
In contrast to the BCS limit, the fast boson exchange results in a non-Fermi-liquid quantum critical state that may be relevant to strange-metal behavior observed in various quantum materials \cite{parkIsotropicQuantumScattering2008,knebelQuantumCriticalPoint2008,doiron-leyraudCorrelationLinearResistivity2009,legrosUniversalTlinearResistivity2019,caoStrangeMetalMagicAngle2020a}.
Quite remarkably, in the presence of both electron and hole Fermi surfaces [see Fig.~\ref{fig:1}], we find a singularity in the $2 k_F$ component of the inter-pocket pair susceptibility at the critical coupling, indicating the onset of PDW order with a spatial oscillation period $\pi/k_F$.
Such PDW correlations can be highlighted by the application of a finite magnetic field, which suppresses the uniform superconducting condensate. 
Under these conditions, the superconducting order manifests as a combination of the inter-pocket $2 k_F$ PDW and slowly-varying intra-pocket superconducting condensate, both of which exhibit the same singularity.

As the microscopic model considered here is not material-specific, we believe that the presented mechanism of the FL instabilities described in this work is relevant for diverse strongly correlated electron systems such as high-temperature superconductors \cite{satoObservationDx2y2LikeSuperconducting2001,doiron-leyraudQuantumOscillationsFermi2007,helmEvolutionFermiSurface2009,jangSuperconductivityInsensitiveOrderEnsuremath2017,liHolePocketDriven2019,fournierInsulatorMetalCrossoverOptimal1998,huckerCompetingChargeSpin2014,satoThermodynamicEvidenceNematic2017}, ultraclean semiconductor quantum wells \cite{mokashiCriticalBehaviorStrongly2012,hossainObservationSpontaneousFerromagnetism2020,melnikovQuantumPhaseTransition2019,shashkinMetalInsulatorTransition2021}, magic-angle twisted bilayer and trilayer graphene (MAG) \cite{caoCorrelatedInsulatorBehaviour2018,sharpeEmergentFerromagnetismThreequarters2019,xieSpectroscopicSignaturesManybody2019,wongCascadeElectronicTransitions2020,jiangChargeOrderBroken2019,luSuperconductorsOrbitalMagnets2019,yankowitzTuningSuperconductivityTwisted2019,caoUnconventionalSuperconductivityMagicangle2018,ohEvidenceUnconventionalSuperconductivity2021,parkTunableStronglyCoupled2021,aroraSuperconductivityMetallicTwisted2020}, 
bilayer transition metal dichalcogenides \cite{wangCorrelatedElectronicPhases2020} and double bilayer graphene \cite{shenCorrelatedStatesTwisted2020,rickhausCorrelatedElectronholeState2021,suSuperconductivityTwistedDouble2023}.

The paper is organized as follows. 
In Sec.~\ref{sec:free} we revisit the long-distance asymptotics of the free (non-interacting) static susceptibilities.
In Sec.~\ref{sec:RG} we provide first- and second-order interaction corrections to static susceptibilities within the leading logarithmic order and construct a one-loop RG description.
The bosonization approach for static susceptibilities is developed in Sec.~\ref{sec:boso}.
Intrinsic instabilities of FL are analyzed in Sec.~\ref{sec:Kohn}.
Applications to different quantum materials and future research directions are discussed in Sec.~\ref{sec:discussion}.
Conclusions are given in Sec.~\ref{sec:conclusions}.
Details of calculations are deferred to Appendices.

\section{Free static susceptibilities}
\label{sec:free}
We consider a zero-temperature, spin-degenerate, $D$-dimensional electron gas, $D > 1$, with two spherical Fermi surfaces, the electron pocket of  radius $k_e$ centered at momentum $\bm K$ and the hole pocket of  radius $k_h$ centered at momentum $-\bm K$, see Fig.~\ref{fig:1}.
If all Fermi surfaces are electron-like (or hole-like), the inter-pocket susceptibilities exhibit the same singularities as the intra-pocket ones.
The electrons interact via a forward-scattering density-density interaction with a finite range $R_0$ that is much greater than the inter-particle distance, i.e. $R_0 \gg \lambda_e = 2\pi/k_e, \lambda_h = 2 \pi/k_h$ [see Eq.~(\ref{inter})].
The long-distance asymptotics of the free electron, $G_e(\tau, \bm r)$, and the free hole, $G_h(\tau, \bm r)$, Green's functions at large distances $r =|{\bm r}| \gg \lambda_e, \lambda_h$, where  $\bm r$ is the $D$-dimensional position coordinate, and large imaginary times $\tau \gg 1/(k_h v_h), 1/(k_e v_e)$, where $v_e > 0$ and $v_h > 0$ are the electron and the hole Fermi velocities, respectively, follows from the general dimensional reduction procedure \cite{miserevDimensionalReductionLuttingerWard2023a},
\begin{eqnarray}
&& G_e(\tau, \bm r) = e^{i \bm K \cdot \bm r} \sum\limits_{\nu_e} \frac{e^{i \nu_e \left(k_e r - \vartheta\right)}}{\left(\lambda_e r\right)^{\frac{D - 1}{2}}} g_{\nu_e} (\tau, r) , \label{Gelectron} \\
&& G_h(\tau, \bm r) = e^{-i \bm K \cdot \bm r} \sum\limits_{\nu_h} \frac{e^{i \nu_h \left(k_h r - \vartheta\right)}}{\left(\lambda_h r\right)^{\frac{D - 1}{2}}} g_{\nu_h} (\tau, r) , \label{Ghole}
\end{eqnarray}
where 
$\nu_e, \nu_h \in \{\pm 1\}$ are the chiral indices, $\vartheta = \pi (D - 1)/4$ is the semiclassical phase, $g_{\nu_e}(\tau, x)$ and $g_{\nu_h}(\tau, x)$ are effective one-dimensional (1D) slowly-varying Green's functions,
\begin{equation}
{\arraycolsep=1.4pt\def\arraystretch{2.2}
\begin{array}{ccl}
\displaystyle g_{\nu_e}(\tau, x) & = & \displaystyle \frac{1}{2\pi} \frac{1}{i \nu_e x - v_e \tau}, \\
\displaystyle g_{\nu_h}(\tau, x) & = & \displaystyle \frac{1}{2\pi} \frac{1}{-i \nu_h x - v_h \tau} . 
\end{array}}
\label{green}
\end{equation}
Note that the hole-like origin of the hole pocket manifests itself in the chirality sign in $g_{\nu_h}(\tau, x)$.

The free static susceptibilities are defined in the standard way,
\begin{eqnarray}
&& \hspace{-10pt} \chi^{(0)}_{P, ab}(\bm r) =  \int\limits_{-\infty}^\infty G_a (\tau, \bm r) G_b(\tau, \bm r) \, d\tau , \\
&& \hspace{-10pt} \chi^{(0)}_{C/S, ab} (\bm r) = -2 \int\limits_{-\infty}^\infty G_a (\tau, \bm r) G_b(-\tau, -\bm r) \, d\tau , \label{spin}
\end{eqnarray}
where $a,b \in \{e, h\}$ and the subscripts $P$, $C$, and $S$ label pair, charge, and spin susceptibilities, respectively.
The free charge and spin susceptibilities are the same for a spin-degenerate electron gas, where the factor of 2 in Eq.~(\ref{spin}) is due to the spin trace.
The intra-pocket pair susceptibilities $\chi^{(0)}_{P, {ee}}(\bm r)$, $\chi^{(0)}_{P, {hh}}(\bm r)$ have a spin-singlet symmetry, while both spin-singlet and spin-triplet symmetry is possible for the inter-pocket pair susceptibility $\chi^{(0)}_{P, {eh}}(\bm r)$.

The intra-pocket pair susceptibility features $\bm Q = \pm 2 \bm K$ long-range harmonics, while the inter-pocket one exhibits the $Q_{eh}^+ = k_e + k_h$ harmonic,
\begin{eqnarray}
&& \chi^{(0)}_{P, {aa}} (\bm r) = \frac{e^{\pm 2 i \bm K \cdot \bm r}}{2 \pi v_a \lambda_a^{D - 1} r^D}, \label{aa}\\
&& \chi^{(0)}_{P, {eh}} (r) = \frac{\cos \left(Q_{eh}^+ r - 2 \vartheta\right)}{\pi (v_e + v_h) \left(\lambda_e \lambda_h\right)^{\frac{D - 1}{2}} r^D} , \label{eh}
\end{eqnarray}
where $a \in \{e, h\}$, and $+2\bm K$ ($-2\bm K$) in Eq.~(\ref{aa}) corresponds to the $ee$ $(hh)$ intra-pocket pair susceptibility.
The intra-pocket pair susceptibilities $\chi^{(0)}_{P, {aa}}(\bm Q)$ probing the zero-momentum pairing, are logarithmically divergent at $\bm Q = \pm 2 \bm K$, where $\chi^{(0)}_{P, {aa}}(\bm Q)$ is the Fourier transform of Eq.~(\ref{aa}) with respect to  ${\bm r}$.
Here, $\bm Q = \pm 2 \bm K$ is due to the choice of origin of the corresponding pocket, see Fig.~\ref{fig:1}.
The Fourier transform $\chi^{(0)}_{P, {eh}}(\bm Q)$ of the inter-pocket pair susceptibility in Eq.~(\ref{eh}) contains a non-analyticity at $|\bm Q| = Q_{eh}^+$ that is not singular at $D > 1$, and oscillates in real space with period $2\pi /Q_{eh}^+$.
The intra-pocket charge and spin susceptibilities exhibit long-range harmonics at $|\bm Q| = 2 k_e$ and $|\bm Q| = 2 k_h$, while the inter-pocket components oscillate at $\bm Q = \pm 2 \bm K + \bm Q_{eh}^-$, $|\bm Q_{eh}^{-}| = Q_{eh}^- = |k_e - k_h|$,
\begin{eqnarray}
&& \chi_{C/S, {aa}}^{(0)}(r) = \frac{\cos \left(2 k_a r - 2 \vartheta\right)}{\pi v_a \lambda_a^{D - 1} r^D}, \\
&& \chi_{C/S, {eh}}^{(0)} (\bm r) = \frac{2 e^{2 i \bm K \cdot \bm r} \cos \left(Q_{eh}^- r\right)}{\pi (v_e + v_h) \left(\lambda_e \lambda_h\right)^{\frac{D - 1}{2}} r^D} ,
\end{eqnarray}
where $a \in \{e, h\}$ and $\chi_{C/S, {he}}^{(0)} (\bm r) = \chi_{C/S, {eh}}^{(0)} (-\bm r)$.
In the fine-tuned case of $k_e = k_h$, the inter-pocket charge and spin susceptibilities $\chi_{C/S, {eh}}^{(0)} (\bm Q) = \chi_{C/S, {he}}^{(0)} (-\bm Q)$ diverge logarithmically in momentum space at $\bm Q = \pm 2 \bm K$.
Otherwise, the Fourier transforms of the free charge and spin susceptibilities are non-singular at $D > 1$.

\begin{figure}[t]
	\centering
	\includegraphics[width=0.99\columnwidth]{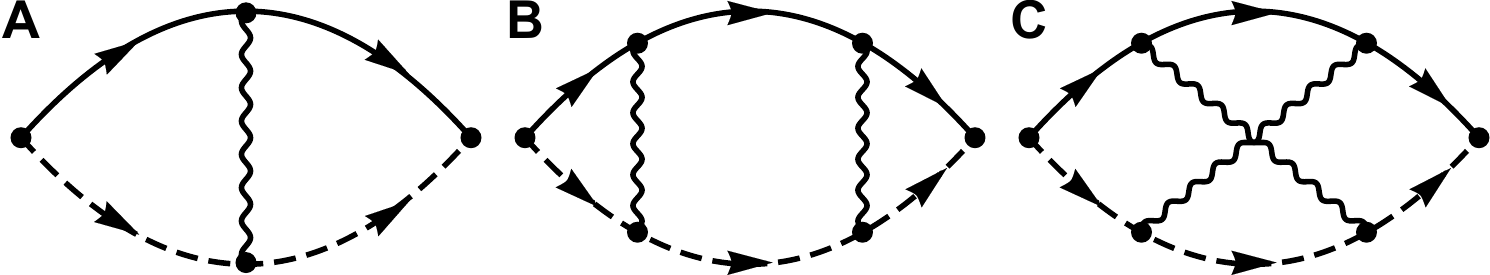}
	\caption{ {\bf Leading logarithmic diagrams for the inter-pocket pair susceptibility in $\mathbf{D > 1}$.}
	A: first-order vertex correction;
	B, C: second-order vertex corrections. 
	The leading logarithmic diagrams for charge and spin susceptibilities share similar structures.
	Solid (dashed) lines stand for the electron (hole) Green's function, while wavy lines denote the  interaction.}
	\label{fig:2}
\end{figure}

\section{Renormalization group}
\label{sec:RG}
Next, we consider interaction corrections to the susceptibilities emerging from a forward-scattering density-density interaction, see Eq.~(\ref{inter}).
Following the recipe of calculation used in Ref.~\cite{hutchinsonSpinSusceptibilityInteracting2023}, which is based on a dimensional reduction technique~\cite{miserevDimensionalReductionLuttingerWard2023a}, we find the first-order interaction corrections to the susceptibilities, see App.~\ref{app:first},
\begin{eqnarray}
&& \chi_{P, {ab}}^{(1)} (\bm r) = - \chi_{P, {ab}}^{(0)}(\bm r) \gamma_{ab} \ln\left(\frac{r}{R_0}\right) , \label{1stPab} \\
&& \chi_{C/S, {ab}}^{(1)}(\bm r) = \chi_{C/S, {ab}}^{(0)}(\bm r) \gamma_{ab} \ln \left(\frac{r}{R_0}\right) , \label{1stCab}
\end{eqnarray}
where $a, b \in \{e, h\}$ and $r \gg R_0$.
Three dimensionless coupling constants are introduced here,
\begin{eqnarray}
&& \gamma_{ee} = \frac{2 V_2}{\pi v_e} , \hspace{3pt} \gamma_{hh} = \frac{2 V_2}{\pi v_h} , \hspace{3pt} \gamma_{eh} = \frac{4 V_2}{\pi (v_e + v_h)} . \label{gamma}
\end{eqnarray}
The notation $V_2$ for the interaction matrix element is taken from Ref.~\cite{hutchinsonSpinSusceptibilityInteracting2023},
\begin{eqnarray}
&& V_2 = \int\limits_0^\infty V(r) \, dr . \label{V2}
\end{eqnarray}
The logarithmic corrections to the susceptibilities originate from vertex corrections, see Fig.~\ref{fig:2}A, and indicate the emergence of singularities.
In case of a repulsive interaction ($\gamma_{ab} > 0$), both inter-pocket and intra-pocket charge and spin susceptibilities are logarithmically enhanced, see Eq.~(\ref{1stCab}), whereas the pair susceptibilities are suppressed, see Eq.~(\ref{1stPab}).
The situation is opposite for an attractive interaction ($\gamma_{ab} < 0$).

In order to construct the one-loop RG, we also require second-order interaction corrections.
Leading second-order diagrams for the inter-pocket pair susceptibility are shown in Fig.~\ref{fig:2}B,C, all other diagrams are subleading in $D \ge 1$, see Apps.~\ref{app:box}, \ref{app:other}, and \ref{app:dynamic}.
Leading logarithmic contributions to charge and spin susceptibilities also originate from the vertex corrections that are diagrammatically similar to the ones shown in Fig.~\ref{fig:2}B,C for the pair susceptibility.
Deferring details of the calculations to App.~\ref{app:box}, we provide here only the final results, accurate to  leading logarithmic order,
\begin{eqnarray}
&& \chi_{P/C/S, {ab}}^{(2)} (\bm r) =  \chi_{P/C/S, {ab}}^{(0)}(\bm r) \frac{\gamma_{ab}^2}{2} \ln^2\!\left(\frac{r}{R_0}\right) , \label{secondorder}
\end{eqnarray}
where, again, $a, b \in \{e, h\}$ and $r \gg R_0$.

The sum of diagrams up to second order for each susceptibility takes the following form:
\begin{eqnarray}
&& \chi(\bm r) = \chi^{(0)}(\bm r) F\left(\frac{r}{R_0}, \gamma_0 \right) , \label{chi} \\
&& F(\rho, \gamma_0) = 1 + \gamma_0 \ln \rho + \frac{\gamma_0^2}{2} \ln^2\!\rho , \label{F}
\end{eqnarray}
where $\gamma_0 = \gamma_{ab}$ for $\chi_{C/S, ab}(\bm r)$ and $\gamma_0 = -\gamma_{ab}$ for $\chi_{P, ab}(\bm r)$.
Here, we follow the multiplicative RG approach \cite{menyhardApplicationRenormalizationGroup1973,solyomApplicationRenormalizationGroup1973}.
Deferring details of the calculations to App.~\ref{app:RG}, we find that the susceptibilities are dressed by the power-law form-factor,
\begin{eqnarray}
&& \chi(\bm r) = \chi^{(0)}(\bm r) \left(\frac{r}{R_0}\right)^{\!{\gamma_0}} . \label{chiRG}
\end{eqnarray}
Here, again, $\gamma_0 = \gamma_{ab}$ for $\chi_{C/S, ab}(\bm r)$ and $\gamma_0 = -\gamma_{ab}$ for $\chi_{P, ab}(\bm r)$.
Note that the exponentiation of the logarithm is also evident from Eq.~(\ref{F}) as it turns out that the interaction coupling constant does not flow within the one-loop RG, see App.~\ref{app:RG}.

\section{Bosonization} 
\label{sec:boso}

In order to support our RG analysis, we apply the bosonization approach \cite{haldaneLuttingerTheoremBosonization2005,castronetoBosonizationFermiLiquids1994,frohlichEffectiveGaugeField1995,metznerFermiSystemsStrong1998,kopietzBosonizationInteractingFermions2006,efetovExactBosonizationInteracting2009,delacretazNonlinearBosonizationFermi2022} that is asymptotically exact in the semiclassical and infrared limit, $r \gg \lambda_e, \lambda_h$ and $\tau \gg 1/(k_h v_h), 1/(k_e v_e)$, respectively~\cite{miserevDimensionalReductionLuttingerWard2023a}.
According to the bosonization \cite{haldaneLuttingerTheoremBosonization2005,castronetoBosonizationFermiLiquids1994,frohlichEffectiveGaugeField1995,metznerFermiSystemsStrong1998,kopietzBosonizationInteractingFermions2006,efetovExactBosonizationInteracting2009,delacretazNonlinearBosonizationFermi2022,miserevDimensionalReductionLuttingerWard2023a}, the random phase approximation (RPA) for the dressed interaction $\tilde V(i \omega, q)$ is asymptotically exact in $D \ge 1$,
\begin{eqnarray}
&& \tilde V(i \omega, q) = \left[V^{-1}(q) - \delta\Pi_0 (i\omega, q)\right]^{-1} , \label{dressedint}
\end{eqnarray}
where $V(q)$ is the $D$-dimensional Fourier transform of the bare interaction $V(r)$, $\delta\Pi_0(i \omega, q) = \Pi_0(i \omega, q) - \Pi_0(0, q \to 0)$ is the dynamic part of the $D$-dimensional polarization bubble $\Pi_0(i \omega, q)$.
As the semiclassical/infrared limit is insensitive to scales $r \sim \lambda_e, \lambda_h$, the analytic part of the exact polarization operator, $\Pi(0, q \to 0) = const.$, must be included in the bare interaction $V(q)$.
For this reason, the matrix element $V_2$ of the forward-scattering interaction is a phenomenological parameter of the problem.

In dimensions $D > 1$ the dynamic screening is irrelevant \cite{chubukovSingularPerturbationTheory2005,gangadharaiahInteractingFermionsTwo2005,aleinerSupersymmetricLowenergyTheory2006,maslovNonanalyticParamagneticResponse2009,zakSpinSusceptibilityInteracting2010,zakFerromagneticOrderNuclear2012}, which is also illustrated in App. \ref{app:dynamic}.
Therefore, we can neglect $\delta \Pi_0(i \omega, q)$ in Eq.~(\ref{dressedint}),
\begin{eqnarray}
&& \tilde V(i \omega, q) \approx V(q), \label{noscreen}
\end{eqnarray}
where $q < 1/R_0$.
In contrast to $D > 1$, in the 1D case the dynamic screening is responsible for the Luttinger-liquid form of the electron Green's function and contributes to $\propto \gamma^2$ corrections to the critical exponents of the 1D susceptibilities \cite{menyhardApplicationRenormalizationGroup1973,solyomApplicationRenormalizationGroup1973,dzyaloshinskiilarkin,delftschoeller,giamarchi}, where $\gamma$ is the interaction coupling constant.
According to Ref.~\cite{miserevDimensionalReductionLuttingerWard2023a}, the dressed interaction $\tilde V(i \omega, r)$, the $D$-dimensional Fourier transform of $\tilde V(i \omega, q)$, plays the role of a dressed 1D interaction, $\tilde V(i \omega, x) = \tilde V_{1\rm D}(i \omega, x)$, which together with Eq.~(\ref{noscreen}) yields,
\begin{eqnarray}
&& \tilde V_{1\rm D} (i \omega, q) \approx \int\limits_{-\infty}^\infty dx \, e^{-i q x}  V(x) \approx 2 V_2 , \label{1Ddressed}
\end{eqnarray}
where $q$ is the 1D momentum transfer, $|q| < 1/R_0$, and $V_2$ is defined in Eq.~(\ref{V2}).
Together with the RPA approximation [see Eq.~(\ref{dressedint})], this allows us to restore an effective 1D bare interaction $V_{1\rm D}(i \omega, q)$, 
\begin{eqnarray}
&& \hspace{-17pt} V_{1\rm D} (i \omega, q) \approx 2 V_2 \left[1 + 2 \sum \limits_{a \in \{e, h\}}  \gamma_{aa} \frac{\omega^2}{\omega^2 + (v_a q)^2}\right]^{-1} \hspace{-3pt} , \label{1Dintapp}
\end{eqnarray}
where $\gamma_{aa}$ is defined in Eq.~(\ref{gamma}), and $v_a$ is the Fermi velocity of pocket $a$.
Here, we used the following form of the dynamic component of the 1D polarization operator,
\begin{eqnarray}
&& \delta\Pi_0^{(1\rm D)} (i\omega, q) = \sum \limits_{a \in \{e, h\}}  \frac{2}{\pi v_a} \frac{\omega^2}{\omega^2 + (v_a q)^2} . \label{pi0}
\end{eqnarray}

In contrast to the conventional Tomonaga-Luttinger model \cite{tomonagaRemarksBlochMethod1950a,luttingerExactlySolubleModel1963,mattisExactSolutionMany1965}, the 1D bare interaction given by Eq.~(\ref{1Dintapp}) is frequency-dependent.
In this paper we aim to calculate critical exponents of $D$-dimensional susceptibilities in linear order in $\gamma_{ab}$ which corresponds to a one-loop RG approximation.
The dynamic part of $V_{1\rm D}(i \omega, q)$ contributes to $\gamma_{ab}^2$ corrections and can therefore be neglected for $\gamma_{ab} \ll 1$,
\begin{eqnarray}
&& V_{1\rm D} (i \omega, q) \approx 2 V_2 . \label{Vapprox}
\end{eqnarray}
The full bosonization treatment accounting for the frequency-dependent part of $V_{1\rm D}(i \omega, q)$ is left for future study.

We stress that the approximation given by Eq.~(\ref{Vapprox}) is valid for correlation functions whose critical exponents scale linearly with $\gamma_{ab}$ for $\gamma_{ab} \ll 1$.
In particular, this approximation is not suitable for fermion Green's functions that require an exact treatment of $V_{1\rm D}(i \omega, q)$.
For this reason, investigation of the fermion Green's function is deferred to future studies.

The general dimensional reduction procedure constructed in Ref.~\cite{miserevDimensionalReductionLuttingerWard2023a} allows us to map $D$-dimensional susceptibilities onto 1D susceptibilities,
\begin{eqnarray}
&& \chi_{P, aa}(\bm r) = \frac{2 e^{\pm 2 i \bm K \cdot \bm r}}{\left(\lambda_a r\right)^{D - 1}} \chi_{P, aa}^{(1\rm D)} (r) , \label{ChiPaadimred} \\
&& \chi_{P, eh} (\bm r) = \frac{2\cos \left(Q_{eh}^+ r - 2 \vartheta \right)}{(\lambda_e \lambda_h)^{\frac{D - 1}{2}} r^{D - 1}} \chi_{P, eh}^{(1\rm D)} (r) , \label{chiPehdimred} \\
&& \chi_{C/S, aa} (\bm r) = \frac{4 \cos \left(2 k_a r - 2 \vartheta\right)}{\left(\lambda_a r\right)^{D - 1}} \chi_{C/S, aa}^{(1\rm D)}(r) , \label{ChiCSaadimred} \\
&& \chi_{C/S, eh} (\bm r) = \frac{4 e^{2 i \bm K \cdot \bm r}\cos \left(Q_{eh}^- r\right)}{(\lambda_e \lambda_h)^{\frac{D - 1}{2}} r^{D - 1}} \chi_{C/S, eh}^{(1\rm D)} (r) , \label{chiCSehdimred} 
\end{eqnarray}
where susceptibilities with the superscript $^{(1\rm D)}$ correspond to conventional 1D susceptibilities, $+2 \bm K$ ($-2 \bm K$) in Eq.~(\ref{ChiPaadimred}) corresponds to $a = e$ ($ a = h$).
We point out that the pocket index plays the role of a pseudo-spin for 1D susceptibilities, all oscillatory factors are taken into account explicitly, see Eqs.~(\ref{ChiPaadimred})--(\ref{chiCSehdimred}).

The bosonization procedure under the approximation given by Eq.~(\ref{Vapprox}) is applied to the 1D susceptibilities and follows a standard 1D bosonization path, e.g., see Refs.~\cite{delftschoeller,giamarchi}, which results in the following scaling behavior,
\begin{eqnarray}
&& \chi_{P, ab} (\bm r) \approx \chi_{P, ab}^{(0)} (\bm r) \left(\frac{r}{R_0}\right)^{\!\!-\gamma_{ab}} , \label{Kpair} \\
&& \chi_{C/S, ab} (\bm r) \approx \chi_{C/S, ab}^{(0)} (\bm r) \left(\frac{r}{R_0}\right)^{\!\gamma_{ab}} , \label{Kcharge}
\end{eqnarray}
where the critical exponents are expanded to linear order in the coupling constants.
The derivation of Eqs.~(\ref{Kpair}) and (\ref{Kcharge}) is given in App.~\ref{app:bosonization}.
As expected, the bosonization approach exactly agrees with the one-loop RG result given by Eq.~(\ref{chiRG}).
We note that the correct treatment of $\propto \gamma_{ab}^2$ and higher-order corrections to the critical exponents would require proper treatment of the dynamic part of the effective 1D bare interaction, see Eq.~(\ref{1Dintapp}). This will be done elsewhere.

We stress that any finite temperature $T > 0$ cuts the power-law singularities at any  coupling strength, see App.~\ref{app:bosonization}.
This behavior is qualitatively different from the BCS-like mean-field mechanisms providing finite critical temperatures at any coupling.
Instead, quite remarkably, the behavior found here is reminiscent of 1D Luttinger liquids, where forward scattering alone is not enough to stabilize  long-range order.
Studies of possible stabilization mechanisms for the relevant orders is deferred to future work.

\begin{figure}[t]
	\centering
	\includegraphics[width=\columnwidth]{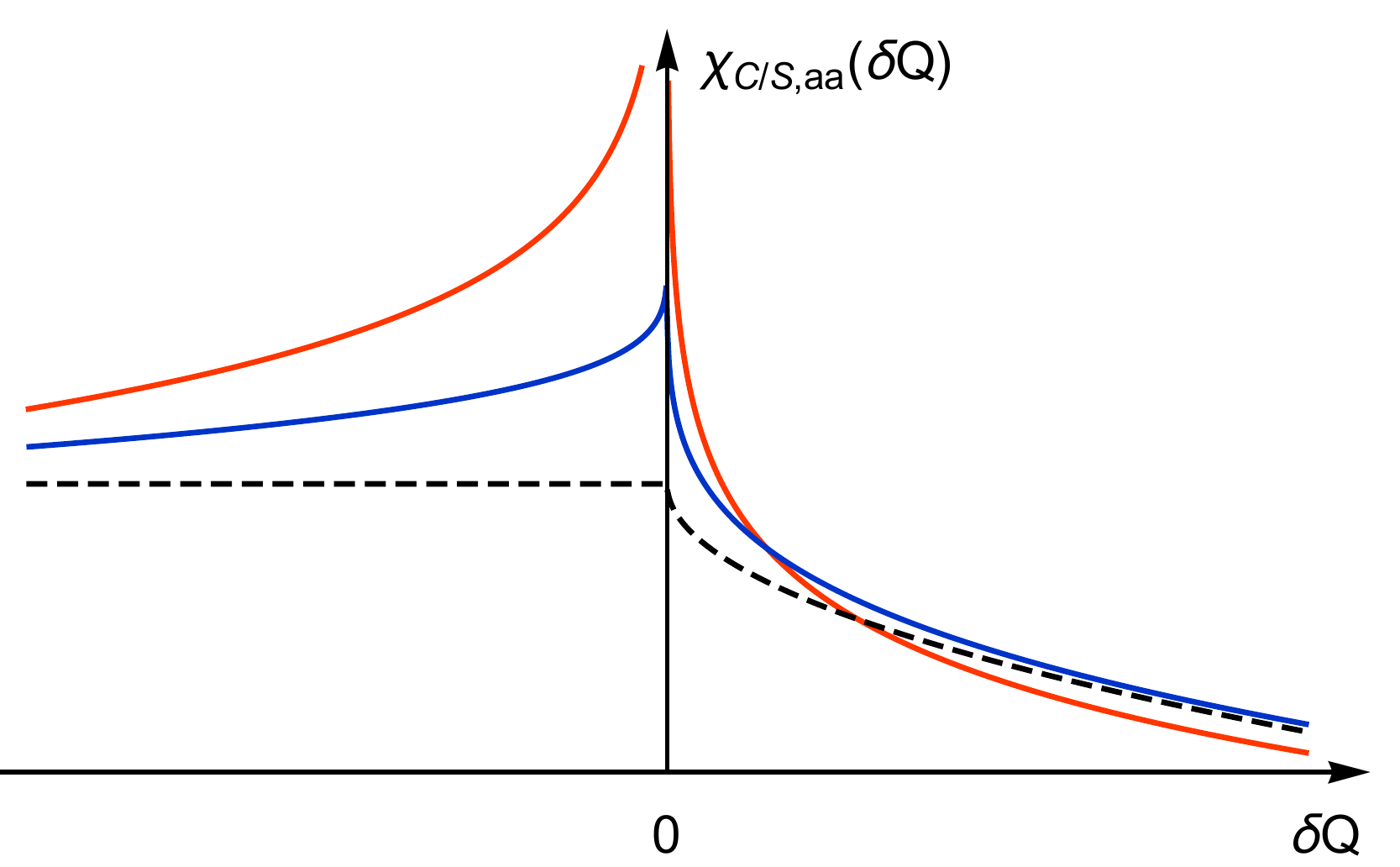}
	\caption{{\bf 2D charge and spin susceptibilities vs. momentum.} 
		Non-analytic part of $\chi_{C/S, aa}(\delta Q)$ as function of $\delta Q = Q - 2 k_a \ll 1/R_0 \ll 2 k_a$ [see Eq.~(\ref{Kohn})], plotted for $D = 2$ at different values of $\gamma_{aa}$:
		1) $\gamma_{aa} = 0$ (black dashed curve) corresponds to the well-known square-root anomaly for the free 2D susceptibilities;
		2) $\gamma_{aa} = 1/4 = \gamma_c /2$ (blue curve) shows a sharp but finite peak near $\delta Q = 0$;
		3) $\gamma_{aa} \to  \gamma_c = 1/2$ (red curve) is logarithmically divergent at $\delta Q = 0$ [see Eq.~(\ref{chargecritaa})], which signals the $Q = 2 k_a$ CDW/SDW instability.
		All three curves are defined up to constant shifts corresponding to the regular analytic parts of the charge and spin susceptibilities at $Q = 2 k_a$.}
	\label{fig:3}
\end{figure}

\section{Instabilities of the Fermi liquid}
\label{sec:Kohn}

In this section we derive the analytical form of the instabilities which signal a breakdown of the Fermi liquid state. First we analyze  repulsive interactions, $\gamma_{ab} > 0$.
In this case only the charge and spin susceptibilities develop relevant singularities, see Eq.~(\ref{Kcharge}).
Taking the $D$-dimensional Fourier transform (see App.~\ref{app:fourier}), we find the singular behavior of these susceptibilities in momentum space,
\begin{eqnarray}
&& \hspace{-17pt} \chi_{C/S, aa}(\delta Q) = \frac{\cos\left[\frac{\pi}{2} \delta\gamma_{aa} + \vartheta \, \mathrm{sgn}(\delta Q)\right] \Gamma\left(\delta\gamma_{aa}\right)}{\pi v_a \left[2 R_0 \lambda_a\right]^{\gamma_c} \left|\delta Q \, R_0 \right|^{\delta\gamma_{aa}}} , \label{Kohn}\\
&& \hspace{-17pt} \chi_{C/S, eh}(\delta Q) \nonumber \\
&& \hspace{17pt} = \frac{2 \cos\left[\frac{\pi}{2} \delta\gamma_{eh} - \vartheta \, \mathrm{sgn}(\delta Q)\right] \Gamma\left(\delta\gamma_{eh}\right)}{\pi (v_e + v_h) \left|R_0 (\lambda_e - \lambda_h)\right|^{\gamma_c} \left|\delta Q \, R_0 \right|^{\delta\gamma_{eh}}} , \label{Kohneh}\\
&& \hspace{-17pt} \gamma_c =\frac{D - 1}{2} , \label{gammac}
\end{eqnarray}
where 
$\delta\gamma_{ab} = \gamma_{ab} - \gamma_c$, $\mathrm{sgn}(x)$ returns the sign of $x$, and $\Gamma(x)$ is the Euler gamma function.
In Eq.~(\ref{Kohn}), $\delta Q = Q - 2 k_a \ll 1/R_0 \ll k_a$, while in Eq.~(\ref{Kohneh}) $\bm Q = 2 \bm K + \bm q$ and $\delta Q = q - Q_{eh}^- \ll \mathrm{min}\{1/R_0, Q_{eh}^-\}$, where $Q_{eh}^- = |k_e - k_h|$.
Here, we point out that Eqs.~(\ref{Kohn}), (\ref{Kohneh}) capture only the non-analytic part of the susceptibilities, the analytic part does not contribute to the long-distance asymptotics given by Eq.~(\ref{Kcharge}).

\begin{figure}[t]
	\centering
	\includegraphics[width=\columnwidth]{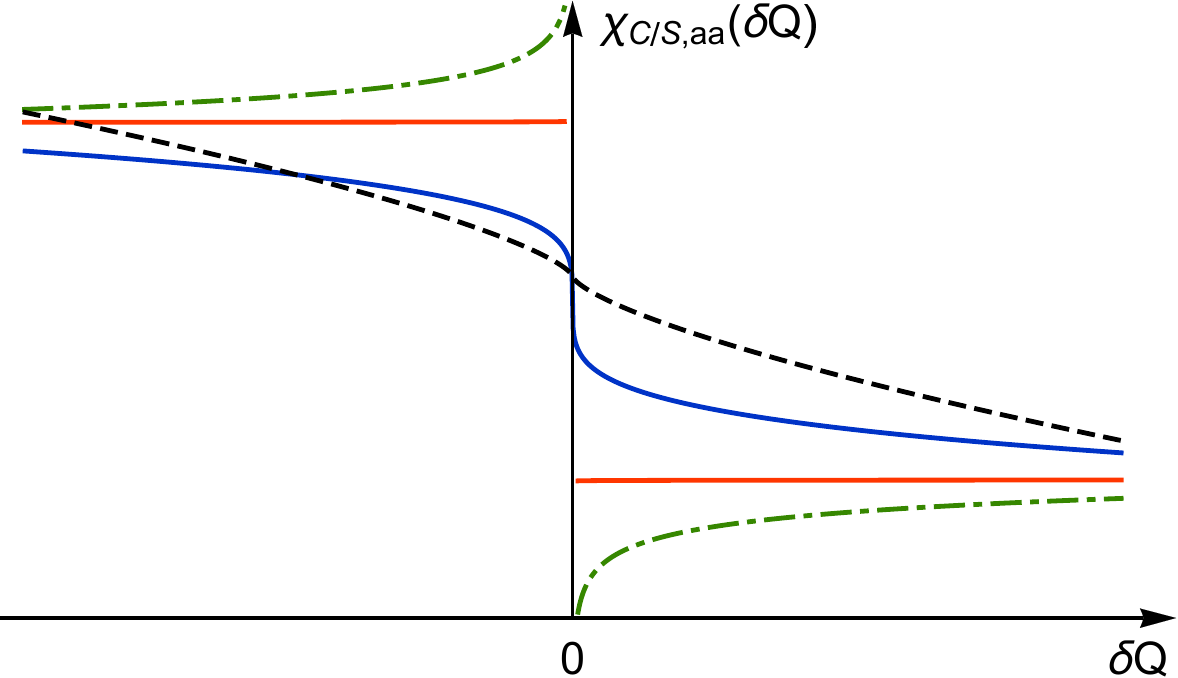}
	\caption{{\bf 3D charge and spin susceptibilities vs. momentum.} 
		Non-analytic part of $\chi_{C/S, aa}(\delta Q)$ as a function of $\delta Q = Q - 2 k_a \ll 1/R_0 \ll 2 k_a$ [see Eq.~(\ref{Kohn})], plotted for $D = 3$ at different values of $\gamma_{aa}$: 
		1) $\gamma_{aa} = 0$ (black dashed curve) corresponds to the conventional anomaly $\propto \delta Q \ln |\delta Q|$ for a free electron gas;
		2) $\gamma_{aa} = 0.75$ (blue curve) shows a much stronger anomaly without a resonant peak at $\delta Q = 0$;
		3) $\gamma_{aa} \to \gamma_c = 1$ (red curve) shows a finite jump at $\delta Q = 0$;
		4) $\gamma_{aa} = 1.1 > \gamma_c$ (dashed dotted green curve) shows a logarithmic singularity at $\delta Q = 0$, signaling the CDW/SDW instability of the 3D interacting electron gas at $\gamma_{ab} > \gamma_c$. All curves are defined up to constant shifts corresponding to the regular analytic parts of the susceptibilities at $Q = 2 k_a$.}
	\label{fig:4}
\end{figure}

For $\gamma_{ab} < \gamma_c$ charge and spin susceptibilities in Eqs.~(\ref{Kohn}), (\ref{Kohneh}) are finite, supporting the FL ground state at weak coupling.
However, at $\gamma_{ab} > \gamma_c$ the susceptibilities diverge at $\delta Q = 0$.
At the quantum critical point, $\delta \gamma_{ab} \to 0$, the critical charge and spin susceptibilities develop the following non-analytic behavior as function of $\delta Q$:
\begin{eqnarray}
&& \hspace{-17pt} \chi_{C/S, aa}(\delta Q) = -\frac{2 \cos \vartheta \ln |\delta Q R_0| + \pi \sin \vartheta \, \mathrm{sgn}(\delta Q)}{2 \pi v_a \left[2 R_0 \lambda_a \right]^{\gamma_c}} , \label{chargecritaa} \\
&& \hspace{-17pt} \chi_{C/S, eh}(\delta Q) = -\frac{2 \cos \vartheta \ln |\delta Q R_0| - \pi \sin \vartheta \, \mathrm{sgn}(\delta Q)}{\pi (v_e + v_h) \left|R_0 \left(\lambda_e - \lambda_h\right) \right|^{\gamma_c}} , \label{chargecriteh}
\end{eqnarray}
where all finite analytic contributions at $\delta Q = 0$ are neglected as they do not lead to singularities at $\delta Q = 0$.
In the two-dimensional (2D) case, $\vartheta = \pi/4$, and the leading singularity is logarithmic with asymmetric peak structure due to the $ \mathrm{sgn}(\delta Q)$-term, see Eqs.~(\ref{chargecritaa}), (\ref{chargecriteh}).
Plots of the 2D charge and spin susceptibilities $\chi_{C/S, aa}(\delta Q)$ for different values of $\gamma_{aa}$ are shown in Fig.~\ref{fig:3}, where all plots are defined up to a constant shift due to regular analytic corrections.

In the three-dimensional (3D) case we get $\vartheta = \pi/2$, which results in a discontinuity $\propto \mathrm{sgn}(\delta Q)$ of the 3D charge and spin susceptibilities, see Fig.~\ref{fig:4}.
For $\gamma_{ab} > \gamma_c$ a logarithmic singularity $\propto \delta \gamma_{ab} \ln |\delta Q R_0| \mathrm{sgn}(\delta Q)$ develops, see green dashed curve in Fig.~\ref{fig:4}.

The singularity in the intra-pocket charge and spin susceptibilities for $\gamma_{aa} > \gamma_c$ signals an instability of the FL towards the formation of  non-uniform insulating CDW/SDW orderings with spatial oscillations with wave vector $2 k_a$.
On the other hand, the singular behavior of the electron-hole (inter-pocket) component of the charge and spin susceptibilities for $\gamma_{eh} > \gamma_c$ signals the formation of an exciton condensate \cite{jeromeExcitonicInsulator1967,khveshchenkoGhostExcitonicInsulator2001,herbutInteractionsPhaseTransitions2006,drutGrapheneVacuumInsulator2009} oscillating in real space with the wave vector $\bm Q_{eh} = 2 \bm K + \bm q_{eh}$, where $q_{eh} = Q_{eh}^-$.
As singularities in both intra- and inter-pocket charge and spin susceptibilities are of the same kind, the ground state is a mixture of $2 k_a$ CDW/SDW and $\bm Q_{eh} = 2 \bm K + \bm q_{eh}$ excitonic orders.

In the fine-tuned case of same-size electron and hole pockets, $k_e = k_h \equiv k_F$, the electron-hole charge and spin susceptibilities $\chi_{C/S, eh}(\bm r) \propto e^{2 i \bm K \cdot \bm r}/r^{D - \gamma_{eh}}$ develop the following power-law singularity (see App.~\ref{app:fourier}):
\begin{eqnarray}
&& \hspace{-15pt}  \chi_{C/S, eh}(q) = \frac{2 \pi^{\frac{D}{2} - 1} \Gamma\left(\frac{\gamma_{eh}}{2}\right)}{\Gamma\left(\frac{D - \gamma_{eh}}{2}\right) (v_e + v_h) \lambda_F^{D - 1}} \left|\frac{2}{q R_0}\right|^{\gamma_{eh}} \! , \label{excfined}
\end{eqnarray}
where $\bm Q = 2 \bm K + \bm q$, $\lambda_F = 2 \pi /k_F$, $q \ll 1/R_0 \ll k_F$.
In this case, the FL is destroyed for arbitrarily small coupling $\gamma_{eh} > 0$ and the ground state is an excitonic insulator \cite{jeromeExcitonicInsulator1967,khveshchenkoGhostExcitonicInsulator2001,herbutInteractionsPhaseTransitions2006,drutGrapheneVacuumInsulator2009}.
The power-law singularity in Eq.~(\ref{excfined}) at $q \to 0$ is due to the instantaneous nature of the Coulomb interaction, see Eq.~(\ref{inter}), which invalidates a standard mean-field approach \cite{jeromeExcitonicInsulator1967,khveshchenkoGhostExcitonicInsulator2001}.

In case of attractive interactions, $\gamma_{ab} < 0$, only pair susceptibilities are relevant, see Eq.~(\ref{Kpair}),
\begin{eqnarray}
&& \hspace{-17pt} \chi_{P, aa}(q) = \frac{\pi^{\frac{D}{2} - 1} \Gamma\left(\frac{|\gamma_{aa}|}{2}\right)}{2 \Gamma\left(\frac{D - |\gamma_{aa}|}{2}\right) v_a \lambda_a^{D - 1}} \left|\frac{2}{q R_0}\right|^{|\gamma_{aa}|} \! , \label{Pairaa} \\
&& \hspace{-17pt} \chi_{P, eh}(\delta Q) = \frac{\cos\left[\frac{\pi}{2} \delta\gamma_{eh} + \vartheta \, \mathrm{sgn}(\delta Q)\right] \Gamma\left(\delta\gamma_{eh}\right)}{\pi (v_e + v_h) \left[R_0 (\lambda_e + \lambda_h)\right]^{\gamma_c} \left|\delta Q \, R_0 \right|^{\delta\gamma_{eh}}} , \label{Paireh}
\end{eqnarray}
where $\delta \gamma_{eh} = |\gamma_{eh}| - \gamma_c$.
In Eq.~(\ref{Pairaa}), $\bm Q = \pm 2 \bm K + \bm q$ and $q \ll 1/R_0 \ll k_a$, where the plus (minus) sign corresponds to $a = e$ ($a = h$).
In Eq.~(\ref{Paireh}), $\delta Q = Q - Q_{eh}^+ \ll 1/R_0 \ll Q_{eh}^+$, where $Q_{eh}^+ = k_e + k_h$.
As follows from Eq.~(\ref{Pairaa}), the FL is destroyed for arbitrarily small $\gamma_{aa} < 0$, and the emerging ground state is a uniform singlet 
($s$-wave) superconductor.
In contrast, a singularity in $\chi_{P, eh}(\delta Q)$ develops only for finite $|\gamma_{eh}| > \gamma_c$, and therefore, this singularity cannot compete with the much stronger power-law singularity in $\chi_{P, aa}(q)$.

In order to highlight the $Q = Q_{eh}^+$ singularity of the electron-hole pair susceptibility, see Eq.~(\ref{Paireh}), we consider a finite Zeeman spin splitting of the Fermi surfaces caused by a magnetic field $\bm B$.
In this case, the intra-pocket pair susceptibility acquires an oscillatory factor $\propto \cos (\Delta k_a r)$, where $\Delta k_a = g_a \mu_B B/v_a$ is the spin splitting of the $a^{\mathrm{th}}$ Fermi surface, $g_a$ is the $g$-factor of the $a^{\mathrm{th}}$ Fermi surface, $\mu_B$ is the Bohr magneton.
A similar oscillatory factor emerges within the Fulde-Ferrel-Larkin-Ovchinnikov (FFLO) mechanism of BCS superconductivity \cite{fuldeSuperconductivityStrongSpinExchange1964,larkinovchinnikov}.
However, in our case the standard mean-field approach \cite{fuldeSuperconductivityStrongSpinExchange1964,larkinovchinnikov} is no longer valid, and the corresponding intra-pocket pair susceptibility develops a non-BCS power-law singularity,
\begin{eqnarray}
&& \chi_{P, aa}(\delta Q) = \frac{\cos\left[\frac{\pi}{2} \delta\gamma_{aa} - \vartheta \, \mathrm{sgn}(\delta Q)\right] \Gamma\left(\delta\gamma_{aa}\right)}{2 \pi v_a \left[2 R_0 \lambda_a\right]^{\gamma_c} \left|\delta Q \, R_0 \right|^{\delta\gamma_{aa}}} , \label{FFLO}
\end{eqnarray}
where $\delta \gamma_{aa} = |\gamma_{aa}| - \gamma_c$, $v_a = (v_a^\uparrow + v_a^\downarrow)/2$ is the average Fermi velocity of the spin-split  pocket $a^{\mathrm{th}}$,  $\lambda_a = (\lambda_a^\uparrow + \lambda_a^\downarrow)/2$ is the average Fermi wavelength,
$\bm Q = \pm 2 \bm K + \bm q$, and $\delta Q = q - \Delta k_a \ll \mathrm{min}\{1/R_0, \Delta k_a\}$.
In this case, both intra-pocket and inter-pocket pair susceptibilities develop the same singularity for $|\gamma_{ab}| > \gamma_c$.
Therefore, the ground state is a mixture of the singlet superconducting order oscillating in real space $\propto e^{\pm 2i \bm K \cdot \bm r} \cos (\Delta k_a r)$ with small wave vector $\Delta k_a = |k_a^\uparrow - k_a^\downarrow|$, and a non-uniform PDW order with a much larger wave vector $|\bm Q_{PDW}| = Q_{eh}^+$.
Here, we point out that the spin-singlet PDW order oscillates with wave vectors $k_e^\uparrow + k_h^\downarrow$, $k_e^\downarrow + k_h^\uparrow$, while the spin-triplet order oscillates with $k_e^\uparrow + k_h^\uparrow$, $k_e^\downarrow + k_h^\downarrow$ wave vectors, which could be used to identify the fraction of spin-triplet superconducting condensate in real materials.
We stress that the PDW order predicted here requires the presence of both electron and hole Fermi surfaces.

Finally, we would like to point out that there exist two dualities that are interchangeable under $\gamma_{ab} \to - \gamma_{ab}$: (i) duality between the intra-pocket charge/spin susceptibilities and the electron-hole pair susceptibility; (ii) duality between the electron-hole charge/spin susceptibilities and the intra-pocket pair susceptibility.
However, we stress that these dualities are no longer exact if higher-order corrections to critical exponents are taken into account.

As our model is not material specific, the behavior uncovered here is universal for all $D$-dimensional metals, $D > 1$, with a finite-range interaction.
We note that effects of strong repulsive interactions can be measured in the FL state for $\gamma_{ab} < \gamma_c$ via local STM measurements of the charge and spin Friedel oscillations decaying via the anomalous power-law $\propto 1/r^{D - \gamma_{ab}}$.
PDW correlations in materials containing both electron and hole pockets can be measured via STM with a superconducting tip \cite{ruanVisualizationPeriodicModulation2018}.

\section{Discussion and future directions}
\label{sec:discussion}

The findings presented here for FL instabilities in dimensions $D > 1$ constitute a solid theoretical ground for unraveling the mechanisms behind quantum phase transitions observed in a variety of quantum materials \cite{agterbergPhysicsPairDensityWaves2020,satoObservationDx2y2LikeSuperconducting2001,doiron-leyraudQuantumOscillationsFermi2007,helmEvolutionFermiSurface2009,jangSuperconductivityInsensitiveOrderEnsuremath2017,liHolePocketDriven2019,fournierInsulatorMetalCrossoverOptimal1998,huckerCompetingChargeSpin2014,satoThermodynamicEvidenceNematic2017,mokashiCriticalBehaviorStrongly2012,hossainObservationSpontaneousFerromagnetism2020,melnikovQuantumPhaseTransition2019,shashkinMetalInsulatorTransition2021,caoCorrelatedInsulatorBehaviour2018,sharpeEmergentFerromagnetismThreequarters2019,xieSpectroscopicSignaturesManybody2019,wongCascadeElectronicTransitions2020,jiangChargeOrderBroken2019,luSuperconductorsOrbitalMagnets2019,yankowitzTuningSuperconductivityTwisted2019,caoUnconventionalSuperconductivityMagicangle2018,ohEvidenceUnconventionalSuperconductivity2021,parkTunableStronglyCoupled2021,aroraSuperconductivityMetallicTwisted2020,wangCorrelatedElectronicPhases2020,shenCorrelatedStatesTwisted2020,rickhausCorrelatedElectronholeState2021,suSuperconductivityTwistedDouble2023}.

In the single-pocket case, we identify  singularities in the $2 k_F$ harmonics of charge and spin susceptibilities for a repulsive interaction at the critical coupling $\gamma = \gamma_c$ (see Eq.~(\ref{Kohn}) and Figs.~\ref{fig:3}, \ref{fig:4}), signaling a phase transition from FL to a strongly correlated CDW/SDW insulator.
We believe that this mechanism of metal-to-insulator transition is relevant to experimental observations in ultraclean semiconductor quantum wells at very low electron densities \cite{mokashiCriticalBehaviorStrongly2012,hossainObservationSpontaneousFerromagnetism2020,melnikovQuantumPhaseTransition2019,shashkinMetalInsulatorTransition2021}.
If the interaction is attractive, a uniform spin-singlet pair susceptibility exhibits a non-BCS power-law singularity at $Q \sim 0$ [see Eq.~(\ref{Pairaa})], indicating a superconducting quantum phase transition.
This mechanism of superconductivity is beyond the standard BCS or Eliashberg treatment based on a mean-filed approach \cite{Eliashberg1960,chubukovEliashbergTheoryPhononmediated2020}.
If the spin degeneracy of the Fermi surface is removed by a finite Zeeman splitting, we find that the leading harmonic in the pair susceptibility is of  FFLO type [see Eq.~(\ref{FFLO})] with a non-BCS singularity emerging at the critical coupling $|\gamma| = \gamma_c$.
Hence,  our theory describes the 
case of non-BCS superconductors which can be relevant for systems with attractive interactions mediated by fast bosons with typical frequency $\omega_0 \gtrsim v_F/ R_0$.
This may correspond to interactions with energetic optical phonons or with collective electron modes (plasmons and spinons)  \cite{takadaPlasmonMechanismSuperconductivity1978,sharmaSuperconductivityCollectiveExcitations2020,grankinInterplayHyperbolicPlasmons2023,abanovQuantumcriticalTheorySpinfermion2003}.
We believe that this mechanism is also relevant to photon-mediated superconductivity in cavity-coupled electron systems \cite{schlawinCavityMediatedElectronPhotonSuperconductivity2019}.

If both electron and hole pockets are present, additional instabilities within the electron-hole channel are possible.
In case of a repulsive interaction, we find that the electron-hole charge and spin susceptibilities develop a singularity at the critical coupling $\gamma_{eh} = \gamma_c$, see Eq.~(\ref{Kohneh}).
If the electron and the hole pockets are of the same size, the singularity develops at arbitrarily small coupling $\gamma_{eh} > 0$, see Eq.~(\ref{excfined}).
This corresponds to a phase transition to an excitonic insulator observed in a variety of layered materials \cite{luZerogapSemiconductorExcitonic2017,maStronglyCorrelatedExcitonic2021,guDipolarExcitonicInsulator2022,jiaEvidenceMonolayerExcitonic2022}.

The two-pocket model with attractive interaction provides a unique mechanism of PDW superconductivity which could be relevant to describe strong electron correlations in cuprate materials~\cite{agterbergPhysicsPairDensityWaves2020,bergCharge4eSuperconductivityPairdensitywave2009,bergStripedSuperconductorsHow2009,agterbergCheckerboardOrderVortex2015,wangPairDensityWaves2018}.
We stress that both electron and hole pockets are required for the PDW superconductivity.
This could be the case for cuprate superconductors due to substantial experimental evidence that points towards a potential Lifshitz transition, resulting in the reconstruction of the Fermi surface into a few small electron and hole pockets near the optimal doping in cuprates \cite{doiron-leyraudQuantumOscillationsFermi2007,helmEvolutionFermiSurface2009,jangSuperconductivityInsensitiveOrderEnsuremath2017,liHolePocketDriven2019}, which lends support to our model.
Attractive interactions in cuprate materials may originate from coupling to high-energy optical phonons.
The finite interaction range $R_0$ in this case has the same electrostatic origin as the Thomas-Fermi screening.
We point out that the phonon energy scale drops out from our model for large phonon frequencies $\omega_0 \gtrsim v_F/R_0$, which corresponds to the limit of instantaneous interaction, see Eq.~(\ref{inter}).
This condition could potentially be satisfied for small electron and hole pockets in cuprate materials.
Therefore, we believe that our model could potentially provide a theoretical basis to describe high-temperature superconductivity in cuprate materials.

The superconducting order observed in MAG and double-bilayer graphene
\cite{luSuperconductorsOrbitalMagnets2019,yankowitzTuningSuperconductivityTwisted2019,caoUnconventionalSuperconductivityMagicangle2018,ohEvidenceUnconventionalSuperconductivity2021,parkTunableStronglyCoupled2021,aroraSuperconductivityMetallicTwisted2020,suSuperconductivityTwistedDouble2023}
is also associated with a Lifshitz transition resulting in reconstructed Fermi surfaces containing both electron and hole pockets \cite{aroraSuperconductivityMetallicTwisted2020,laksonoSingletSuperconductivityEnhanced2018,liuChiralSpinDensity2018}.
This could potentially signal the PDW origin of superconductivity in these strongly correlated quantum materials as well. 
It might be possible to change the interaction range $R_0$ in these materials in a controlled way by gates, which affect the screening, and thereby test our predictions.

The mechanism of FL instabilities uncovered here is not material-specific but rather universal for interacting electron systems in dimensions $D > 1$.
Therefore, our model and  theory could also be relevant to experimental observations of quantum phase transitions in other materials \cite{wangCorrelatedElectronicPhases2020,shenCorrelatedStatesTwisted2020,rickhausCorrelatedElectronholeState2021,suSuperconductivityTwistedDouble2023}.

Further  directions of study encompass 
bosonization techniques to calculate higher-order corrections to the critical exponents that are required to evaluate the electron spectral function.
Quantum critical behavior of finite-temperature susceptibilities when the interaction coupling is greater than the critical value, points to strange-metal physics which is typical for quantum critical systems \cite{parkIsotropicQuantumScattering2008,knebelQuantumCriticalPoint2008,doiron-leyraudCorrelationLinearResistivity2009,legrosUniversalTlinearResistivity2019,caoStrangeMetalMagicAngle2020a}.
A detailed theoretical description of these effects is currently under way and left for a subsequent study.
Mechanisms stabilizing long-range orders at finite temperatures will also be considered elsewhere.

\section{Conclusions}
\label{sec:conclusions}
We proposed and analyzed a realistic microscopic model universally describing superconducting and insulating instabilities of interacting $D$-dimensional Fermi liquids observed in a large variety of quantum materials, from ultraclean semiconductor quantum wells at very low electron densities to high-temperature superconductors.
In case of a single Fermi surface, we predict CDW/SDW instabilities at the critical value of repulsive interaction coupling and non-BCS singlet superconductivity for an attractive interaction.
In case if both electron and hole pockets are present, we find an instability towards excitonic insulators if the interaction is repulsive, while towards $k_e + k_h$ PDW superconductivity if the  interaction is attractive with a coupling strength  that is greater than a critical universal value.
Stabilization mechanisms of long-range orders at finite temperature, as well as their melting and possible formation of the strange-metal state at high temperatures is deferred to  subsequent studies.

\begin{acknowledgments}
We wish to acknowledge hospitality at The  Anthony J. Leggett Institute for Condensed Matter Theory, Urbana (USA), during the inauguration workshop New Horizons in Condensed Matter Physics, November, 2023, which stimulated this work.
We also thank Dr. Valerii Kozin for valuable comments.
This work was supported by the Georg H. Endress Foundation and the Swiss National Science Foundation (SNSF).
This project has received funding from the European Union’s Horizon 2020 research and innovation programme under Grant Agreement No. 862046.
\end{acknowledgments}

\appendix

\section{First-order diagrams}
\label{app:first}
In the Appendices, we provide technical details on the first- and second-order susceptibility diagram calculations, renormalization group construction, $D$-dimensional Fourier transforms, and bosonization.
As the calculations for all susceptibility diagrams are quite similar, we provide here the details for the inter-pocket pair susceptibility only.
In order to calculate the Feynman diagrams, we employ the dimensional reduction technique \cite{miserevDimensionalReductionLuttingerWard2023a}, which greatly simplifies the evaluations since $D$-dimensional integrals are reduced to effective 1D integrations (with some exceptions, see below).
Dimension-reduced representations for the susceptibility diagrams can be derived in full analogy with the general derivations presented in Ref.~\cite{miserevDimensionalReductionLuttingerWard2023a}.
For example, the long-distance asymptotics of the static inter-pocket pair susceptibility takes the following form, [see also Eq.~(\ref{chiPehdimred})]
\begin{eqnarray}
&& \hspace{-12pt} \chi_{P, eh} (\bm r) = \sum\limits_{\nu_e, \nu_h} \frac{e^{i r (\nu_e k_e + \nu_h k_h) - i \vartheta (\nu_e + \nu_h)}}{(\lambda_e \lambda_h)^{\frac{D - 1}{2}} r^{D - 1}} \chi_{\nu_e \nu_h} (r) , \label{chiP}
\end{eqnarray}
where $\nu_e, \nu_h \in \{\pm 1\}$ are the chiral indices, $\vartheta = \pi (D - 1)/4$ is the semiclassical phase, the function $\chi_{\nu_e \nu_h} (r)$ is represented by the dimension-reduced susceptibility diagrams, the index $^{(1\rm D)}$ introduced in Eq.~(\ref{chiPehdimred}) is suppressed here for brevity.
In Eq.~(\ref{chiPehdimred}), it is already taken into account that only the $\nu_e = \nu_h$ components exhibit non-analytic behavior.
We show this directly in Apps. \ref{app:first}, \ref{app:box}.
We emphasize that polarization effects must be treated within the original $D$ dimensions \cite{miserevDimensionalReductionLuttingerWard2023a}.
This is important for the dynamic screening contribution in the second-order diagrams considered below. 
More complicated multiloop diagrams do not appear within the first two lowest orders \cite{miserevDimensionalReductionLuttingerWard2023a}, except the second-order Aslamazov-Larkin diagrams for charge and spin susceptibilities, see Fig.~\ref{SMfig:2}B, which, however, vanish in electron gas with forward-scattering density-density interaction.

The first-order diagram contributing to the leading logarithmic order corresponds to the vertex correction, see Fig.~\ref{fig:2}A.
The required calculations are similar to the ones shown in Ref.~\cite{hutchinsonSpinSusceptibilityInteracting2023}, so we only provide a few key steps here.
The leading logarithmic first-order correction to 
$\chi_{\nu_e \nu_h} (r)$ is represented by the 1D analog of the vertex-correction diagram in Fig.~\ref{fig:2}A,
\begin{eqnarray}
&& \hspace{-15pt} \chi^{(1)}_{\nu_e \nu_h} (r) = -\int d\tau \int d\xi_1 d\xi_2 \, U(\xi_1 - \xi_2) g_{\nu_e} (\xi - \xi_1) \nonumber \\
&& \hspace{23pt} \times g_{\nu_e}(\xi_1) g_{\nu_h}(\xi - \xi_2) g_{\nu_h} (\xi_2) , \label{vertex1}
\end{eqnarray}	
where $\xi = (\tau, r)$ and $\xi_i = (\tau_i, x_i)$. Here, $\tau_i$ is the imaginary time, $x_i \in \mathbb{R}$ is the 1D spatial coordinate, $d\xi_i = d\tau_i dx_i$, $U(\xi) = U(\tau, |x|) = \delta(\tau) V(|x|)$ is the electron-electron interaction, and $g_{\nu_e}(\xi)$ and $g_{\nu_h}(\xi)$ are the 1D electron and hole Green's function, respectively [see Eq.~(\ref{green})].	
As we are interested in the asymptotics at large distances, $r \gg R_0 \gg \lambda_e, \lambda_h$, the interaction $U(\xi_1 - \xi_2)$ is effectively local on this scale (while still being non-local on the scale of the inter-particle distance) and we can approximate $\xi_2 \approx \xi_1$ in Eq.~(\ref{vertex1}),
\begin{eqnarray}
&& \hspace{-15pt} \chi_{\nu_e \nu_h}^{(1)}(r) = - 2 V_2 \int d\tau \int d\xi_1 \, \nonumber \\
&& \hspace{23pt} \times g_{\nu_e} (\xi - \xi_1)  g_{\nu_e}(\xi_1) g_{\nu_h}(\xi - \xi_1) g_{\nu_h} (\xi_1) , \label{vertex2}
\end{eqnarray}
where $V_2$ is given by Eq.~(\ref{V2}).
We point out that the dimensional reduction is suitable for all interactions with the range $R_0 \gg \lambda_e, \lambda_h$.
This procedure can not be applied to  zero-range contact interactions.
Integration over $\tau$ immediately shows that only terms with $\nu_e = \nu_h$ contribute.
Otherwise, the poles of $g_{\nu_e} (\xi - \xi_1)$ and $g_{\nu_h} (\xi - \xi_1)$ for $\nu_e \neq \nu_h$ are located on the same complex half-plane and the integral over $\tau$ vanishes.
After performing elementary contour integrations over $\tau$ and $\tau_1$, we find,
\begin{eqnarray}
&& \chi_{\nu_e \nu_h}^{(1)}(r) = -\frac{\gamma_{eh} \delta_{\nu_e, \nu_h}}{8 \pi v_+} \int \frac{dx_1}{|x_1| |x_1 - r|} ,
\end{eqnarray}
where $v_+ = v_e + v_h$, $\gamma_{eh}$ is the dimensionless coupling constant [see Eq.~(\ref{gamma})], and $\delta_{\nu_e, \nu_h}$ is the Kronecker symbol.
The integral over $x_1$ diverges at short distances $|x_1| \lesssim R_0$, $|x_1 - r| \lesssim R_0$ because we neglected details of the interaction on short scales $\sim R_0$.
This ultraviolet divergence is therefore cured by restricting the integration to the interval $x_1 \in (-\infty, -R_0) \cup (R_0, r - R_0) \cup (r + R_0, +\infty)$, where $r \gg R_0$,
\begin{eqnarray}
&& \int \frac{dx_1}{|x_1| |x_1 - r|} \approx \frac{4}{r} \ln\left|\frac{r}{R_0}\right| . \label{int1}
\end{eqnarray}
Substituting this result back into Eq.~(\ref{vertex1}), we find,
\begin{eqnarray}
&& \chi_{P, eh}^{(1)}(\bm r) = -\chi_{P, eh}^{(0)}(r) \gamma_{eh} \ln\left|\frac{r}{R_0}\right| .
\end{eqnarray}
Again, this asymptotics is valid at $r \gg R_0 \gg \lambda_e, \lambda_h$.

The first-order corrections to the intra-pocket charge and spin susceptibilities calculated in Ref.~\cite{hutchinsonSpinSusceptibilityInteracting2023} are dual to the corresponding corrections to the inter-pocket pair susceptibility upon the transformation $v_h \to v_e$, $\nu_h \to - \nu_e$ and $\tau \to -\tau$ in the hole Green's function. 
The same transformation reveals the duality between the inter-pocket charge and spin susceptibilities and the intra-pocket pair susceptibility.

\section{Box and cross-box diagrams}
\label{app:box}
The leading second-order diagrams, see Fig.~\ref{fig:2}B,C, correspond to insertion of the box and cross-box scattering amplitude diagrams into the inter-pocket pair susceptibility.
The same insertions into the charge and spin susceptibilities also produce leading logarithmic second-order diagrams that are calculated similarly to the pair susceptibility box and cross-box diagrams shown in Fig.~\ref{fig:2}B,C.
Corresponding contributions to $\chi_{\nu_e \nu_h}(r)$, see Eq.~(\ref{chiP}), are represented by 1D diagrams.
Applying the locality condition for large-distance correlations at $r \gg R_0$, we find,
\begin{eqnarray}
&& \hspace{-10pt} \chi_{\nu_e \nu_h}^{(b)} (r) = 4 V_2^2 \int d\tau \int d\xi_1 d\xi_2 \, g_{\nu_e} (\xi - \xi_1) g_{\nu_e} (\xi_1 - \xi_2) \nonumber \\
&& \hspace{28pt} \times g_{\nu_e} (\xi_2) g_{\nu_h} (\xi -\xi_1) g_{\nu_h} (\xi_1 - \xi_2) g_{\nu_h}(\xi_2) , \label{box1} \\
&& \hspace{-10pt} \chi_{\nu_e \nu_h}^{(cb)} (r) = 4 V_2^2 \int d\tau \int d\xi_1 d\xi_2 \, g_{\nu_e} (\xi - \xi_1) g_{\nu_e} (\xi_1 - \xi_2) \nonumber \\
&& \hspace{28pt} \times  g_{\nu_e} (\xi_2) g_{\nu_h} (\xi -\xi_2) g_{\nu_h} (\xi_2 - \xi_1) g_{\nu_h}(\xi_1) , \label{cbox1}
\end{eqnarray}
where, again, $\xi = (\tau, r)$, $\xi_i = (\tau_i, x_i)$, and the superscripts $^{(b)}$ and $^{(cb)}$ stand for the box and cross-box diagram, respectively.

The box diagram can be integrated in a fashion that is similar to the first-order vertex correction.
First, we integrate over $\tau$ and find that terms only with $\nu_e = \nu_h$ contribute. 
After integrating over $\tau$, $\tau_1$, and $\tau_2$, which are simple contour integrals, we find,
\begin{eqnarray}
&& \hspace{-10pt} \chi_{\nu_e \nu_h}^{(b)}(r) = \frac{\gamma_{eh}^2 \delta_{\nu_e, \nu_h}}{32 \pi v_+} \int \frac{dx_1 dx_2}{|x_2| |x_2 - x_1| |x_1 - r|} , \label{box2}
\end{eqnarray}
where, again, we have to cut out the interval $(-R_0, R_0)$ near each short-distance singularity in Eq.~(\ref{box2}).
The integral over $x_1$ and $x_2$ can be calculated via a partial fraction expansion of the integrand, yielding
\begin{eqnarray}
&& \int \frac{dx_1 dx_2}{|x_2| |x_2 - x_1| |x_1 - r|} = \frac{12}{r} \ln^2 \left|\frac{r}{R_0}\right| , \label{int2}
\end{eqnarray}
where $r \gg R_0$, and we keep only the $\log^2$ contribution.
Therefore, the box diagram gives the following contribution,
\begin{eqnarray}
&& \chi_{P, eh}^{(b)} (\bm r) = \frac{3}{4} \gamma_{eh}^2 \ln^2\left|\frac{r}{R_0}\right| \chi_{P, eh}^{(0)} (r) , \label{box}
\end{eqnarray}
where $r \gg R_0$.

The cross-box diagram in Fig.~\ref{fig:2}C is more complicated, see Eq.~(\ref{cbox1}).
In order to integrate over $\xi_2$, we separate products of 1D Green's functions into sums, for example,
\begin{eqnarray}
&& \hspace{-20pt} g_{\nu_e}(\xi_1 - \xi_2) g_{\nu_e} (\xi_2) \nonumber \\
&& \hspace{45pt} = g_{\nu_e} \left(\xi_1\right) \left[g_{\nu_e} \left(\xi_1 - \xi_2\right) + g_{\nu_e} \left(\xi_2\right)\right] , \label{gg}
\end{eqnarray}
and similar for hole Green's functions.
Then, the integral over $\xi_2$ is especially straightforward,
\begin{eqnarray}
&& \hspace{0pt} \chi_{\nu_e \nu_h}^{(cb)} (r) = -2 V_2 \int d\tau \int d\xi_1 \, g_{\nu_e} \left(\xi - \xi_1\right) g_{\nu_e} \left(\xi_1\right) \nonumber \\
&& \hspace{38pt} \times  g_{\nu_h} \left(\xi - \xi_1\right) g_{\nu_h} \left(\xi_1\right) \mathcal{I}_{\nu_e \nu_h}(\xi_1, \xi) , \label{cbox2} \\
&& \hspace{0pt} \mathcal{I}_+(\xi_1, \xi) = \frac{\gamma_{eh}}{4} \left\{\ln \left[\left(\zeta_{h} - \zeta_{1h}\right)\left(\zeta_{e} - \zeta_{1e} \right) \zeta_{1e} \zeta_{1h} \frac{v_e v_h}{v_+^2 R_0^4}\right] \right. \nonumber \\
&& \hspace{38pt} \left. - \ln\left[\frac{\zeta_{e} \zeta_h}{R_0^2}\right] \right\} , \label{Iplus} \\
&& \hspace{0pt} \mathcal{I}_-(\xi_1, \xi) = \frac{V_2}{\pi (v_e - v_h)} \ln \left[\frac{v_h \zeta_h}{v_e \zeta_e} \frac{\left(\zeta_{e} - \zeta_{1e}\right) \zeta_{1e}}{\left(\zeta_{h} - \zeta_{1h}\right) \zeta_{1h}}\right] , \label{Iminus} 
\end{eqnarray}
where we introduced the following complex variables,
\begin{eqnarray}
&& \zeta_e = x + i \nu_e v_e \tau, \hspace{5pt} \zeta_h = x - i \nu_h v_h \tau , \label{zetaeh} \\
&& \zeta_{1e} = x_1 + i \nu_e v_e \tau_1, \hspace{5pt} \zeta_{1h} = x_1 - i \nu_h v_h \tau_1 . \label{zetaeh1}
\end{eqnarray}
Here, $\mathcal{I}_+(\xi_1, \xi)$ [$\mathcal{I}_-(\xi_1, \xi)$] corresponds to $\mathcal{I}_{\nu_e \nu_h}(\xi_1, \xi)$ at $\nu_e = \nu_h$ ($\nu_e = - \nu_h$). 
Note that Eq.~(\ref{cbox2}) has the same structure as the first-order vertex correction, see Eq.~(\ref{vertex2}), but being multiplied by the factor $\mathcal{I}_{\nu_e \nu_h}(\xi_1, \xi)$.
As here we are interested in the leading logarithmic order, terms of order unity are not shown in $\mathcal{I}_{\nu_e, \nu_h}(\xi_1, \xi)$.

In case of $\nu_e = -\nu_h$, the integral over $\mathcal{I}_-(\xi_1, \xi)$ vanishes.
This is best seen if $\mathcal{I}_-(\xi_1, \xi)$ is expanded in a sum of logarithms and then the integrations are performed for each logarithmic term separately.
For example, the integration with $\ln (\zeta_{e} - \zeta_{1e})$ gives zero after the shift $\xi_1 \to \xi - \xi_1$ after which the logarithm transforms into $\ln (\zeta_{1e})$ which is independent of $\tau$.
The integral over $\tau$ is then taken over $g_{\nu_e}(\xi - \xi_1) g_{\nu_h} (\xi - \xi_1)$ and vanishes if $\nu_e = -\nu_h$.
The same is true for the other logarithmic terms in Eq.~(\ref{Iminus}).

In case of $\nu_e = \nu_h$, it is also convenient to expand $\mathcal{I}_+(\xi_1, \xi)$ into a sum of logarithmic terms, and then perform appropriate coordinate shifts to transform Eq.~(\ref{cbox2}) to the following form,
\begin{eqnarray}
&& \hspace{-20pt} \chi_{\nu_e \nu_h}^{(cb)}(r) = \delta_{\nu_e, \nu_h} \left(\chi_1^{(cb)}(r) + \chi_2^{(cb)}(r) \right), \\
&& \hspace{-20pt} \chi_1^{(cb)}(r) = \frac{-\gamma_{eh}^2}{32 \pi^2} \! \int \! \frac{d \tau_1 dx_1}{\zeta_{1e} \zeta_{1h}} \ln\left(\! \frac{v_e v_h \zeta_{1e} \zeta_{1h}}{v_+^2 R_0^2}\! \right) \! \frac{1}{|x_1 - r|} , \label{chi1} \\
&&\hspace{-20pt} \chi_2^{(cb)}(r) = \frac{\gamma_{eh}^2}{32 \pi^2} \int \frac{d\tau}{\zeta_{e} \zeta_{h}} \ln^2 \left(\frac{\zeta_e \zeta_h}{R_0^2}\right) , \label{chi2}
\end{eqnarray}
where $v_+ = v_e + v_h$.
Here, we separated parts that contain logarithms depending on $\zeta_{1e}$ and $\zeta_{1h}$ [first term in Eq.~(\ref{Iplus})] from the ones that only depend on $\zeta_e$ and $\zeta_h$ [second term in Eq.~(\ref{Iplus})].
As the logarithms in $\chi_1^{(cb)}(r)$ are independent of $\tau$, the integration over $\tau$ was also performed in Eq.~(\ref{chi1}).
As the logarithm-terms in $\chi^{(cb)}_2(r)$ are independent of $\xi_1$, this allowed us to perform the integration,
\begin{eqnarray}
&& \hspace{-20pt}  \int \frac{d\xi_1}{\zeta_{1e} \zeta_{1h}(\zeta_{1e} - \zeta_e) (\zeta_{1h} -\zeta_h)} = \frac{4 \pi \ln \left(\zeta_e \zeta_h/R_0^2\right)}{\zeta_e \zeta_h v_+} .
\end{eqnarray}
The result is shown in Eq.~(\ref{chi2}).
Integration over $\tau_1$ in Eq.~(\ref{chi1}) is transformed into a contour integral with logarithmic singularity which is then evaluated by elementary integration,
\begin{eqnarray}
&& \int \frac{d \tau_1}{2 \pi} \frac{\ln (\zeta_{1e} \zeta_{1h})}{\zeta_{1e} \zeta_{1h}} = \frac{1}{|x_1| v_+} \ln\left(\frac{x_1^2 v_+^2}{v_e v_h}\right) .
\end{eqnarray}
The remaining integral over $x_1$ was calculated before, see Eqs.~(\ref{int1}), (\ref{int2}), 
after which we find the simple relation
\begin{eqnarray}
&& \chi_1^{(cb)}(r) = -\chi^{(b)}(r) ,
\end{eqnarray}
where $\chi^{(b)}(r)$ is given by Eq.~(\ref{box2}) for $\nu_e = \nu_h$.
Therefore, the remaining contribution $\chi_2^{(cb)}(r)$ represents the sum of box and cross-box diagrams.
The integral over $\tau$ in Eq.~(\ref{chi2}) is elementary, and within the leading logarithmic order we obtain,
\begin{eqnarray}
&& \chi_2^{(cb)}(r) = \frac{\gamma_{eh}^2}{4 \pi v_+ r} \ln^2\left|\frac{r}{R_0}\right| .
\end{eqnarray}
Therefore, we find that the sum of the two diagrams in Fig.~\ref{fig:2}B, C results in Eq.~(\ref{secondorder}).

Applying the transformation $v_h \to v_e$, $\nu_h \to - \nu_e$ and $\tau \to - \tau$, we obtain the corresponding diagrams for the intra-pocket charge and spin susceptibilities.

\section{Subleading second-order diagrams}
\label{app:other}

Here we discuss the remaining second-order diagrams except the dynamic screening diagrams [see Fig.~\ref{SMfig:3}A,B] which are considered separately in App.~\ref{app:dynamic}.

\begin{figure}[t]
	\centering
	\includegraphics[width=\columnwidth]{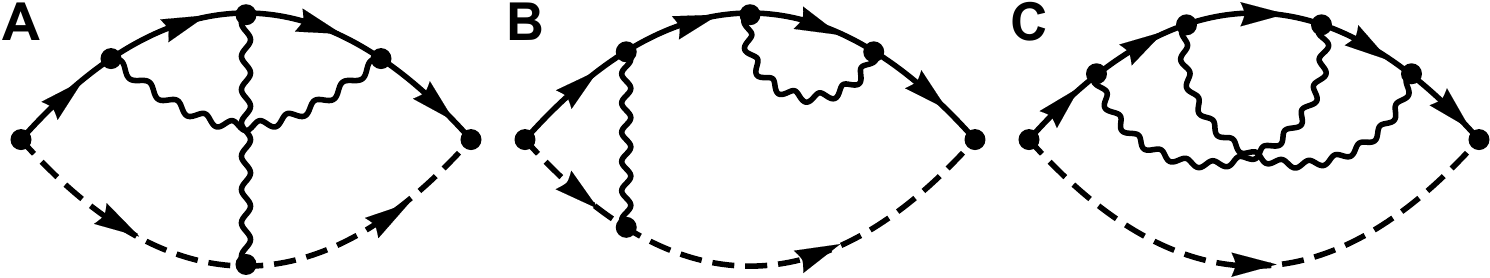}
	\caption{{\bf Subleading diagrams.}
		Diagrams containing dressing of the interaction vertex (A), dressing of Green's functions by the Fock self-energy (B), and dressing of Green's functions by the crossed self-energy diagram (C).
		In all diagrams, full (dashed) directed lines denote the electron (hole) Green's function, while wavy lines stand for the interaction.}
	\label{SMfig:1}
\end{figure}

First, we consider the diagram in Fig.~\ref{SMfig:1}A containing dressing of the interaction vertex,
\begin{eqnarray}
&& \hspace{-10pt} \chi_{\nu_e \nu_h}^{(v)} (r) = 4 V_2^2  \int  d\tau \int  d\xi_1 d\xi_2 \, g_{\nu_e} (\xi - \xi_1) g_{\nu_e}(\xi_1 - \xi_2) \nonumber \\
&& \hspace{28pt} \times g_{\nu_e}(\xi_2 - \xi_1) g_{\nu_e}(\xi_1) g_{\nu_h}(\xi - \xi_2) g_{\nu_h}(\xi_2) , \label{chiv}
\end{eqnarray}
where the superscript $^{(v)}$ stands for the interaction vertex and $\xi = (\tau, r)$.
Applying Eq.~(\ref{gg}) and using $g_{\nu_e} (-\xi) = - g_{\nu_e}(\xi)$, we simplify this expression to the following form,
\begin{eqnarray}
&& \hspace{-10pt} \chi_{\nu_e \nu_h}^{(v)} (r) = -8 V_2^2 \int d\tau \, g_{\nu_e}(\xi) g_{\nu_h}(\xi) \int  d\xi_1 d\xi_2 \, g_{\nu_e}(\xi_1) \nonumber \\
&& \hspace{28pt} \times \left[g_{\nu_e}(\xi_1 - \xi_2)\right]^2 \left[g_{\nu_h}(\xi_2) +  g_{\nu_h}(\xi - \xi_2)\right] .
\end{eqnarray}
Integration over $\xi_1$ is elementary,
\begin{eqnarray}
&& \int d\xi_1 \, g_{\nu_e}(\xi_1) \left[g_{\nu_e}(\xi_1 - \xi_2)\right]^2 = \frac{\tau_2}{4 \pi^2 \zeta_{2e}^2} ,
\end{eqnarray}
where $\zeta_{2e} = x_2 + i \nu_e v_e \tau_2$.
The remaining integrals over $\xi_2$ and $\tau$ are also elementary contour integrals, so we find 
\begin{eqnarray}
&& \chi^{(v)}_{\nu_e \nu_h}(r) = -\frac{\gamma_{eh}^2 \delta_{\nu_e \nu_h}}{8 \pi v_+ r} \ln \left(\frac{r v_+}{\sqrt{v_e v_h}}\right) .
\end{eqnarray}
Indeed, we see that the diagram in Fig.~\ref{SMfig:1}A gives a subleading contribution that can be safely neglected.

\begin{figure}[t]
	\centering
	\includegraphics[width=0.8\columnwidth]{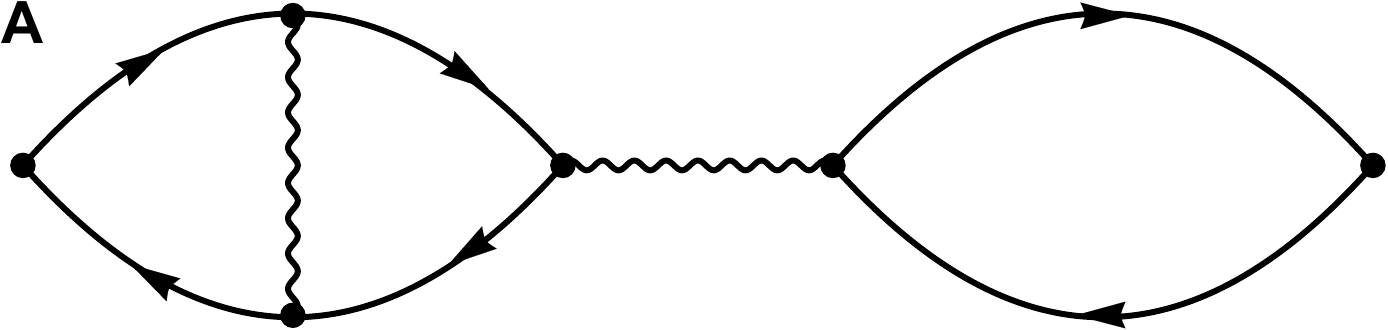}\\
	\vspace{10pt}
	\includegraphics[width=0.6\columnwidth]{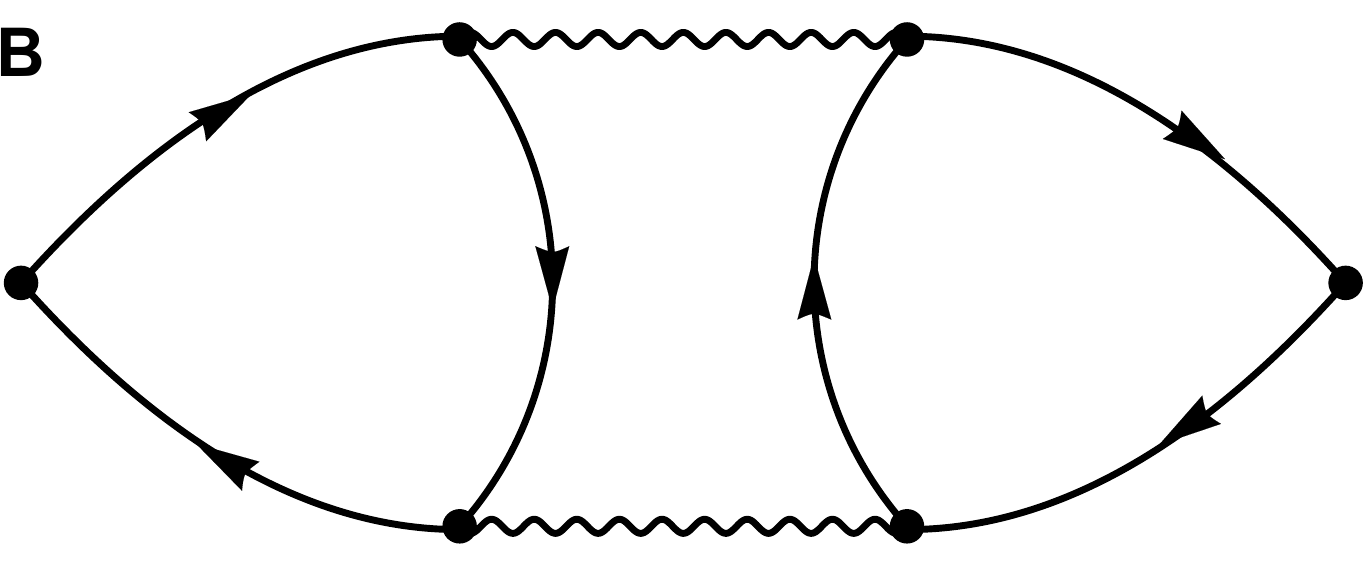}
	\caption{{\bf Additional second-order diagrams for charge and spin susceptibilities.}
		These diagrams do not appear for the pair susceptibility. 
		Diagram B is known as the Aslamazov-Larkin diagram.
		The second Aslamazov-Larkin diagram (not shown here) corresponds to the reversal of Green's function arrows in one of the loops.}
	\label{SMfig:2}
\end{figure}

Diagrams containing dressing of Green's functions by the Fock self-energy, e.g. the diagram in Fig.~\ref{SMfig:1}B, are also subleading because the first-order self-energy is irrelevant at $r \gg R_0$.

Next, we calculate the diagram in Fig.~\ref{SMfig:1}C that contains dressing of the electron Green's function by the crossed self-energy diagram,
\begin{eqnarray}
&& \hspace{-15pt} \chi^{(se)}_{\nu_e \nu_h}(r) = - 4 V_2^2 \int d\tau \, g_{\nu_h} (\xi) \int d\xi_1 d\xi_2 \,    \nonumber \\
&& \hspace{23pt} \times g_{\nu_e}(\xi - \xi_1) \left[g_{\nu_e}(\xi_1 - \xi_2)\right]^3 g_{\nu_e} (\xi_2) , \label{chise}
\end{eqnarray}
where $\xi = (\tau, r)$.
After straightforward integration we find no logarithms,
\begin{eqnarray}
&& \chi^{(se)}_{\nu_e \nu_h}(r) = -\frac{\gamma_{eh}^2 \delta_{\nu_e, \nu_h}}{64 \pi v_+ r} ,
\end{eqnarray}
and thus this contribution is also negligible.

There are also diagrams for charge and spin susceptibilities shown in Fig.~\ref{SMfig:2}A,B.
All these diagrams are clearly subleading.
The Aslamazov-Larkin diagram shown in Fig.~\ref{SMfig:2}B vanishes for the spin susceptibility due to zero spin trace.
For the charge susceptibility, the sum of the two Aslamazov-Larkin diagrams (the second diagram differs from the one shown in Fig.~\ref{SMfig:2}B by reversal of Green's function arrows in one of the loops) vanishes for the $Q = 0$ harmonic of the charge susceptibility due to destructive interference between the two diagrams.
The Aslamazov-Larkin diagrams do not contribute to the finite-momentum anomaly of the charge susceptibility due to  momentum conservation: the forward scattering interaction transfers small momenta $Q \lesssim 1/R_0 \ll k_e, k_h$ and cannot obtain a large $2 k_F$ momentum from a fermion loop.

\section{Dynamic screening}
\label{app:dynamic}
\begin{figure}[t]
	\centering
	\includegraphics[width=\columnwidth]{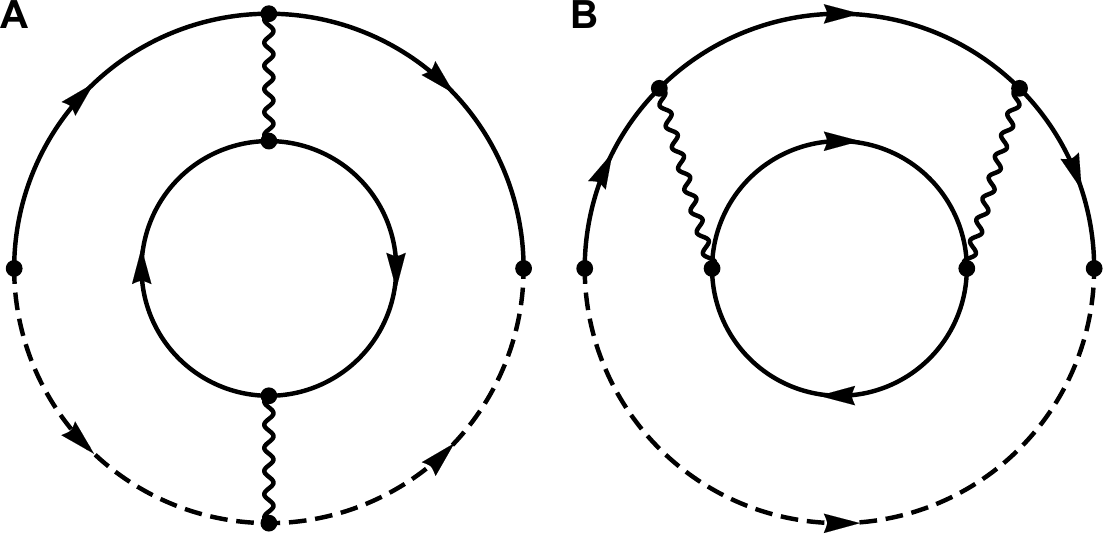}
	\caption{{\bf Dynamic screening diagrams.} Here, only the dressing of the interaction line by the electron particle-hole bubble is shown, dressing by the hole bubble is treated similarly.}
	\label{SMfig:3}
\end{figure}

Dynamic screening diagrams for the pair susceptibility are shown in Fig.~\ref{SMfig:3}A,B.
These diagrams cannot be mapped onto its 1D analogue straightforwardly \cite{miserevDimensionalReductionLuttingerWard2023a}.
Here, we show details for the diagram in Fig.~\ref{SMfig:3}A.
Conclusions about the diagram in Fig.~\ref{SMfig:3}B follow from similar reasoning.

According to the dimensional reduction procedure \cite{miserevDimensionalReductionLuttingerWard2023a}, the dynamic screening contribution to $\chi_{\nu_e \nu_h}(r)$ [see Eq.~(\ref{chiP})] can be represented as follows,
\begin{eqnarray}
&& \hspace{-15pt} \chi_{\nu_e \nu_h}^{(scr)} (r) = - \int d\tau \int d\xi_1 d\xi_2 \, \delta U (\xi_1 - \xi_2) \nonumber \\
&& \hspace{24pt} \times g_{\nu_e}(\xi - \xi_1)  g_{\nu_e}(\xi_1) g_{\nu_h}(\xi - \xi_2) g_{\nu_h} (\xi_2) , \label{scr}
\end{eqnarray}
where $\delta U(\xi) = \delta U(\tau, r)$ is given by the $D$-dimensional diagram representing the dressing of the interaction by the polarization operator $\Pi(\tau, r)$,
\begin{eqnarray}
&& \delta U (z) = \int dz_1 dz_2 \, U(z - z_1) \Pi(z_1 - z_2) U(z_2) , \label{dU}
\end{eqnarray}
where $z = (\tau, \bm r)$, $z_i = (\tau_i, \bm r_i)$ and $\bm r, \bm r_i$ are $D$-dimensional coordinates.
As the bare interaction is of finite-range forward scattering type and we are interested in the physics at large distances $r \gg R_0$, 
Eq.~(\ref{dU}) can be simplified to the following form,
\begin{eqnarray}
&& \delta U(\tau, r) \approx V_0^2 \, \Pi(\tau, r) , \label{V0} \\
&& \Pi(\tau, r) = 2 \sum\limits_{a} G_a(\tau, \bm r) G_a(-\tau, -\bm r) , \label{Pi} 
\end{eqnarray}
where $a \in \{e, h\}$, $V_0 = \int V(r) \, d\bm r = V(Q = 0)$ is the Fourier transform of the bare interaction at small momentum transfers (we take the limit $Q = 0$). 
The polarization operator is given by the particle-hole bubble, its long-distance asymptotic behavior follows from the asymptotics of the Green's functions [see Eqs.~(\ref{Gelectron}), (\ref{Ghole})],
\begin{eqnarray}
&& \hspace{-35pt} \Pi(\tau, r) = 2 \sum\limits_{a} \sum\limits_{\nu_{1a}, \nu_{2a}} \frac{e^{i (\nu_{1a} - \nu_{2a}) (k_a r - \vartheta)}}{(\lambda_a r)^{D - 1}} \nonumber \\
&& \hspace{-2pt} \times g_{\nu_{1a}}(\tau, r) g_{\nu_{2a}}(-\tau, - r) , \label{Piasympt}
\end{eqnarray}
where $a \in \{e, h\}$, $\nu_{1a}, \nu_{2a} \in \{\pm 1\}$ are chiral indices, see also Ref.~\cite{miserevDimensionalReductionLuttingerWard2023a}.

Finite-momentum anomalies in $\Pi(\tau, r)$ contribute to the $2 k_e$ and $2 k_h$ harmonics of the interaction that does not contribute to the finite-momentum anomaly of the pair susceptibility in the diagram in Fig.~\ref{SMfig:3}A due to momentum conservation.
Therefore, only the $\nu_{1a} = \nu_{2a}$ term in Eq.~(\ref{Piasympt}) contributes to $\chi^{(scr)}_{\nu_e \nu_h}(r)$ at $\nu_e = \nu_h$, see Eq.~(\ref{scr}),
\begin{eqnarray}
&& \Pi^{q\sim 0} (\tau, r) = 2 \sum\limits_{a, \nu_a} \frac{g_{\nu_a} (\tau, r) g_{\nu_a} (-\tau, - r)}{(\lambda_a r)^{D - 1}} , \label{Pi0}
\end{eqnarray}
where $a \in \{e, h\}$, $\nu_a \in \{\pm 1\}$.
Using Eq.~(\ref{gg}), we transform Eq.~(\ref{scr}) into the following form,
\begin{eqnarray}
&& \hspace{0pt} \chi_+^{(scr)} (r) = -\sum\limits_{a, \nu_a} \! \int \! \frac{V_0^2 \, d\tau}{\pi^2 \lambda_a^{D - 1}} g_{\nu_e}(\xi) g_{\nu_h}(\xi) \!  \int \! \frac{d\xi_1 d\xi_2}{|x_1 - x_2|^{D - 1}} \nonumber\\
&& \hspace{39pt} \times \frac{g_{\nu_e} (\xi_1)}{(\zeta_{1a} - \zeta_{2a})^2} \left[g_{\nu_h}(\xi_2) + g_{\nu_h}(\xi - \xi_2)\right] , \label{scr2}
\end{eqnarray}
where the subscript $_+$ indicates that we consider the case $\nu_e = \nu_h$.
After performing integrations over $\tau$, $\tau_1$, and $\tau_2$, which are all elementary contour integrals, we find,
\begin{widetext}
\begin{eqnarray}
&& \hspace{-17pt} \chi_+^{(scr)}(r) = - \frac{V_0^2}{2 \pi^3 v_e v_h v_+ r} \sum\limits_{a}\frac{1}{v_a^2 \lambda_a^{D-1}} \! \int\limits_0^r \!\! \int\limits_0^\infty \!\! \frac{d x_1 dx_2}{|x_1 - x_2|^{D - 1}} \! \left\{\! \left[\frac{x_1}{v_e} + \frac{x_2}{v_h} +\frac{|x_1 - x_2|}{v_a}\right]^{-2} \!\!\! +  \left[\frac{x_1}{v_h} + \frac{x_2}{v_e} +\frac{|x_1 - x_2|}{v_a}\right]^{-2} \! \right\} .\label{scr3}
\end{eqnarray}
\end{widetext}
It does not matter which integration interval corresponds to which variable, due to the symmetry between $x_1$ and $x_2$. 
The factor $1/|x_1 - x_2|^{D - 1}$ in Eq.~(\ref{scr3}) is responsible for a dramatic suppression of this diagram in $D > 1$, which scales $\propto 1/r^D$.
In contrast, all other diagrams for $\chi_{\nu_e\nu_h}(r)$ which we calculated before scale as $1/r$ modulo powers of logarithms.
The $\propto 1/r^D$ scaling of $\chi_+^{(scr)}(r)$ is evident after rescaling $x_1 \to r u_1$, $x_2 \to r u_2$, the remaining integral over $u_1$ and $u_2$ is convergent after appropriate regularization of the kernel $1/|x_1 - x_2|^{D - 1}$.
The ultraviolet divergence near $x_1 = x_2$ originates from the semiclassical asymptotics of the Green's function [see Eqs.~(\ref{Gelectron}), (\ref{Ghole})] and cannot be applied if $|x_1 - x_2| \lesssim \lambda_e, \lambda_h$. 
The diagram in Fig.~\ref{SMfig:3}B is irrelevant in $D > 1$ for the same reason.
We note that the diagram in Fig.~\ref{SMfig:3}B is subleading even if $D = 1$ (not considered here) and does not affect susceptibilities within one-loop RG~\cite{menyhardApplicationRenormalizationGroup1973,solyomApplicationRenormalizationGroup1973}.

\section{Renormalization group}
\label{app:RG}
In this section we derive the RG flow equation for the coupling constant following a multiplicative RG procedure.
We begin with the observation that the susceptibilities take the following form,
\begin{eqnarray}
&& \chi(r) = \chi^{(0)}(r) F\left(\frac{r}{R_0}, \gamma_0\right) ,
\end{eqnarray}
where, again, $\chi^{(0)}(r)$ is the free susceptibility, $R_0$ the finite range of the interaction, and $\gamma_0$ the  interaction coupling.
However, from a theoretical point of view, we could choose a different value of the cut-off, say $R_0'$, yet we expect the form-factor $F$ to retain the same dependence on $r$, just with a new value of the cut-off $R'_0$ and, possibly, new value of the coupling constant.
This observation corresponds to the following scaling symmetry of the form-factor~\cite{menyhardApplicationRenormalizationGroup1973,solyomApplicationRenormalizationGroup1973}:
\begin{eqnarray}
&& F\left(\frac{r}{R_0}, \gamma(R_0)\right) = Z F\left(\frac{r}{R_0'}, \gamma \left(R_0'\right)\right), \label{RGscale}
\end{eqnarray}
where $\gamma(R_0')$ is the coupling associated with the new cut-off $R_0'$ and $\gamma (R_0) = \gamma_0$ is our original (bare) coupling.
The multiplicative factor $Z$ does not depend on $r$.
We can get rid of this factor by taking the logarithm of Eq.~(\ref{RGscale})
and differentiate it by $r$,
\begin{eqnarray}
&& \hspace{-10pt} r \frac{\partial}{\partial r} \ln F\left(\frac{r}{R_0}, \gamma \left(R_0\right)\right) = r \frac{\partial}{\partial r} \ln F\left(\frac{r}{R_0'}, \gamma \left(R_0'\right)\right) .
\end{eqnarray}
Note that $R_0$ and $R_0'$ are independent parameters.
In particular, we can choose $R_0' \to r$ which establishes the flow equation for the form-factor,
\begin{eqnarray}
&& \frac{\partial \ln F\left(\rho, \gamma \left(R_0\right)\right)}{\partial \ln \rho}  = \lim\limits_{y \to 1} \frac{\partial \ln F\left(y, \gamma(r)\right)}{\partial \ln y} , \label{flow}
\end{eqnarray} 
where we introduced the notations $y = r/R_0'$, $\rho = r/R_0$ is the dimensionless coordinate, and the limit $R_0' = r$ is taken.
Note that the right-hand side of Eq.~(\ref{flow}) just gives us the first coefficient in the Taylor series expansion of $\ln F (y, \gamma)$ with respect to $\ln y$.
This coefficient can be calculated within the perturbation theory that provides us with this Taylor series expansion.
For example, for the pair susceptibility we find,
\begin{eqnarray}
&& \hspace{-15pt} \frac{\chi_{P, eh}(r)}{\chi_{P, eh}^{(0)}(r)} \approx  1 - \gamma_{eh} \ln \left|\frac{r}{R_0}\right| + \frac{\gamma_{eh}^2}{2} \ln^2 \left|\frac{r}{R_0}\right| + \dots \label{ourF}
\end{eqnarray}
In particular, this means that to the linear order in $\ln y$ the logarithmic form-factor in our case has the following expansion,
\begin{eqnarray}
&& \ln F(y, \gamma) = 1 + \gamma \ln y + \frac{\beta(\gamma)}{2}\ln^2 y + \dots , \label{lnF}
\end{eqnarray}
where $\beta(\gamma)$ is the RG $\beta$-function that we discuss in a moment.
Therefore, it is convenient to introduce the coupling constant such that it straight away corresponds to the coefficient of the $\ln y$ term in $\ln F (y, \gamma)$, i.e. the flow equation for the form-factor in our case takes the form,
\begin{eqnarray}
&& \frac{\partial \ln F\left(\rho, \gamma \left(R_0\right)\right)}{\partial \ell}  = \gamma(r) , \label{flow2}
\end{eqnarray}
where $\rho = r/R_0$, $\ell = \ln \rho$.
We point out that for the inter-pocket pair susceptibility the coupling constant $\gamma = - \gamma_{eh}$.
The formal dependence of the coupling constant on the coordinate $r$ in Eq.~(\ref{flow2}) is due to the limit $R_0' \to r$.
The coupling constant is a function of the cut-off  (not of the coordinate $r$).

In order to obtain the flow equation for the coupling constant, we just act on both sides of Eq.~(\ref{RGscale}) by the operator $\partial^2/\partial \ell^2$, and use the same limit $R_0' = r$ to obtain
\begin{eqnarray}
&&\hspace{-17pt} \frac{\partial^2 \ln F\left(\rho, \gamma \left(R_0\right)\right)}{\partial \ell^2} = \lim\limits_{y \to 1} \frac{\partial^2 \ln F(y, \gamma(r))}{\partial (\ln y)^2} = \beta(\gamma(r)) .
\end{eqnarray}
Using the definition of our coupling constant, we find the flow equation,
\begin{eqnarray}
&& \frac{\partial \gamma(R_0')}{\partial \ell} = \beta(\gamma) , \label{B}
\end{eqnarray}
where $\ell = \ln (R_0'/R_0)$.
Here, we emphasize again that the coupling constant is a function of the cut-off $R_0'$ and not of the coordinate $r$.

The Taylor expansion of the pair susceptibility form-factor given by Eq.~(\ref{ourF}) yields us a vanishing $\beta$-function, $\beta(\gamma) = 0$, in leading order.
This means that the coupling does not flow, i.e. $\gamma(R_0') = \gamma(R_0)$.
The only reason why we get $\beta(\gamma) = 0$ is because we were only taking into account terms $\propto \gamma^2 \ell^2$.
However,  $\ell^2$ terms are also produced in higher orders of the perturbation expansion, for example, in third-order diagrams we can have $\gamma^3 \ell^2$ contributions.
This means that here we have proven the following relation:
\begin{eqnarray}
&& \frac{\partial \gamma(R_0')}{\partial \ell} = \mathcal{O}\left(\gamma^3\right) . \label{cube}
\end{eqnarray}
In order to make a conclusive statement on how the coupling constant runs, a two-loop RG is required, i.e. one has to consider all third-order diagrams with accuracy $\gamma^3 \ell^2$ which is generally challenging (and not considered here).

Here we present the leading-order one-loop RG behavior.
In this case we find that the coupling constant does not flow which immediately solves the flow equation for the susceptibility form-factor,
\begin{eqnarray}
&& F\left(\frac{r}{R_0}, \gamma\right) = \left(\frac{r}{R_0}\right)^{\! \gamma} .
\end{eqnarray}
This establishes one of the main results of this work [see Eq.~(\ref{chiRG})] allowing us to draw significant conclusions on the instabilities in an interacting electron gas in $D$ dimensions.

We also would like to demonstrate the importance of the cross-box diagram shown in Fig.~\ref{fig:2}C in the present context.
For this, we consider the susceptibility form-factor in absence of this cross-box diagram  and obtain,
\begin{eqnarray}
&& F_{Cooper}(\rho, \gamma) = 1 + \gamma \ln \rho + \frac{3}{4} \gamma^2 \ln^2 \rho +\dots ,
\end{eqnarray}
where the subscript ``Cooper'' indicates that only the Cooper ladder diagrams, see Fig.~\ref{fig:2}B, are retained in the susceptibility via the RG procedure.
This corresponds to $\beta(\gamma) = \gamma^2/2$ with the Cooper-pole singularity of the running coupling constant,
\begin{eqnarray}
&& \gamma_{Cooper}(R_0') = \frac{\gamma_0}{1 - \frac{\gamma_0}{2} \ell}\, ,
\end{eqnarray}
where $\ell = \ln (R_0'/R_0)$.
In turn, this results in the double-pole structure of the susceptibility form-factor,
\begin{eqnarray}
&& F_{Cooper}(\rho, \gamma) = \frac{1}{\left[1 - \frac{\gamma_0}{2} \ell\right]^2} \, ,
\end{eqnarray}
where $\ell = \ln \rho$, $\rho = r/R_0$.
Here, $\gamma_0 = -\gamma_{ab}$ for the pair susceptibility and $\gamma_0 = \gamma_{ab}$ for charge and spin susceptibilities.
Presence of the pole in the susceptibility form-factor provides a finite energy scale corresponding to a finite gap at any $\gamma_0 > 0$.
The cross-box diagram completely removes the pole and yields the power-law singularity that corresponds to quantum critical behavior discussed in the main text.

\section{Bosonization}
\label{app:bosonization}
	
In this Appendix we show that a standard bosonization calculation in strict 1D gives the same result for the scaling of the susceptibilities in leading order as the RG approach. In this calculation we neglect the dynamic part of the effective 1D bare interaction [see Eq.~(\ref{1Dintapp})] since it plays no role for the scaling in leading order, as was shown in App. \ref{app:dynamic}. 
	
The effective 1D susceptibilites introduced in Eqs.~(\ref{ChiPaadimred}-\ref{chiCSehdimred}) can be expressed as
\begin{align}
  \nonumber
  \chi^{(1\rm D)}_{C/S,ab}(x) & \\
  &\hspace{-1cm}
  = \int d\tau \langle T (\psi_{+a\uparrow} \psi_{-b\uparrow}^\dagger)(\xi) 
	(\psi_{+a\uparrow}\psi_{-b\uparrow}^\dagger)^\dagger (0)\rangle , \label{C_susceptibility} \\
  \nonumber 
  \chi^{(1\rm D)}_{P,ab}(x) & \\
  &\hspace{-1cm}
    = \int d\tau \langle T (\psi_{+a\uparrow} \psi_{-b\uparrow})(\xi)
    (\psi_{+a\uparrow}\psi_{-b\uparrow})^\dagger (0)\rangle \,, \label{P_susceptibility}
\end{align}
where $\xi = (\tau, x)$, $T$ is the time-ordering operator, and, as before, $a=b \in \{e,h\}$ ($a=e$, $b=h$) correspond to the intra-pocket (inter-pocket) case. 
The fermion field operator $\psi_{\nu a\sigma}(x)$ corresponds to the slowly varying part of the right ($\nu = +$) and left ($\nu=-$) movers with band index $a \in \{e, h\}$ and spin index $\sigma \in \{\uparrow, \downarrow\}$. 
Due to the spin degeneracy, we choose all fermion fields with spin $\sigma = \uparrow$, other cases yield the same result. 
We also use the convention that the chirality index of holes is opposite to the right/left moving index, see Eq.~(\ref{green}). 
	
Using conventional bosonization \cite{delftschoeller,giamarchi}, we express slowly varying right/left moving field operators in the following form (the Klein factors are discarded for simplicity)
\begin{align}
		\label{psi_rl}
		\psi_{\nu a\sigma}(x) & = \frac{1}{\sqrt{2\pi R_0}} \,e^{i\nu\sqrt{4\pi}\varphi_{\nu a\sigma}(x)} \,,\\
		\nonumber
		\varphi_{\nu a\sigma}(x) & = \sum\limits_{q\ne 0} \left(\frac{1}{2L|q|}\right)^{1/2} \theta(\nu q)\\ 
		\label{bosonic_fields}
		&\times \left(e^{iqx}a_{q,a\sigma} + e^{-iqx}a_{q,a\sigma}^\dagger\right)e^{-R_0 |q|/2}\,.
\end{align}
Here, $R_0$ is the short distance cutoff given by the range of the forward scattering, $L$ is the length of the system, $q=2\pi n/L$ are the discrete momenta for periodic boundary conditions, and $a_{q,a\sigma}$, $a_{q,a\sigma}^\dagger$ denote bosonic operators. 
	
The bosonized Hamiltonian with forward scattering only reads
\begin{eqnarray}
	&& \hspace{-20pt} H = \sum\limits_{\nu a\sigma} v_a \int dx \, \left(\partial_x \varphi_{\nu a\sigma}\right)^2  \nonumber \\
	&& + \frac{V_2}{\pi}\int dx \, \left(\sum_{\nu a\sigma}\partial_x\varphi_{\nu a \sigma}\right)^2\,.
	\label{H_bosonized}
\end{eqnarray}
Introducing canonically conjugate fields $\varphi_{a\sigma}=\varphi_{+a\sigma} + \varphi_{-a\sigma}$ and $\Pi_{a\sigma}=-\partial_x\vartheta_{a\sigma}$, where $\vartheta_{a\sigma}=\varphi_{+a\sigma} - \varphi_{-a\sigma}$, together with the charge and spin fields $\varphi^C_a=\frac{1}{\sqrt{2}}\sum_\sigma\varphi_{a\sigma}$, $\varphi^S_a=\frac{1}{\sqrt{2}}\sum_\sigma\sigma\varphi_{a\sigma}$ (and analogously for $\vartheta^{C/S}_a$ and $\Pi^{C/S}_a$), we find the spin-charge separated form for $H$:
\begin{eqnarray}
	&& \hspace{-15pt} H = H_C + H_S , \label{H_CS} \\
	&& \hspace{-15pt} H_C = \sum\limits_a \frac{v_a}{2} \int dx \, \left[\left(\partial_x \varphi_a^C\right)^2 + \left(\Pi_a^C\right)^2\right] \nonumber \\
	&& \hspace{15pt} + \frac{2V_2}{\pi}\int dx \, \left(\sum\limits_a \partial_x\varphi_a^C\right)^2 \,,  \label{H_C}\\
	&& \hspace{-15pt} H_S = \sum\limits_a\frac{v_a}{2} \int dx \, \left[\left(\partial_x \varphi_a^S\right)^2 + \left(\Pi_a^S\right)^2\right]
		\,. 	\label{H_S}
\end{eqnarray}
The spin part is independent of the interaction and contributes only to the free part of the susceptibilities. Therefore, we consider in the following only the charge part.
	
After a straightforward diagonalization, we can write the charge part in the following form of free bosons with linear dispersion relation,
\begin{align}
	\label{H_diagonalized}
	H_C &= \sum\limits_a\frac{\tilde{v}_a}{2} \int dx \, \left[\left(\partial_x \tilde{\varphi}_a\right)^2 + \left(\tilde{\Pi}_a\right)^2\right]
		\,,
\end{align}
where $\tilde{v}_a=\bar{v}/g_a$ are renormalized velocities and the new canonically conjugate fields $\tilde{\varphi}_a$ and $\tilde{\Pi}_a=-\partial_x\tilde{\vartheta}_a$ are defined as follows:
\begin{align}
		\label{new_varphi_fields}
		\tilde{\varphi}_a &= \sum\limits_b \frac{1}{\sqrt{g_a}} \,U_{ab} \,\frac{1}{\sqrt{\alpha_b}} \,\varphi^C_b \,,\\
		\label{new_vartheta_fields}
		\tilde{\vartheta}_a &= \sum\limits_b \sqrt{g_a} \,U_{ab} \,\sqrt{\alpha_b} \,\vartheta^C_b \,.
\end{align}
Here, the following notations are introduced,
\begin{eqnarray}
	&& \alpha_a = \frac{v_a}{\bar{v}} , \hspace{5pt} \bar{v}=\frac{1}{2}\left(v_e + v_h\right) \,, \label{alpha_vbar}\\
	&& g_a = \left(\alpha^2 + 2\gamma_{eh} + a\lambda\right)^{-1/2} \,, \label{g}\\
	&& \lambda = \sqrt{\delta\alpha^2 \left(1+\gamma_{eh}\right) + 4\alpha_+\alpha_-\gamma_{eh}^2}\,, \label{lambda}
\end{eqnarray}
where $\alpha^2=\frac{1}{2}(\alpha_+^2 + \alpha_-^2)$, $\delta\alpha=\alpha_+-\alpha_-$. The unitary matrix $U$ diagonalizes the matrix $A$ defined as follows,
\begin{align}
		\nonumber
		A &= (\alpha^2 + 2\gamma_{eh})\left(\begin{array}{cc} 1 & 0 \\ 0 & 1\end{array}\right) + \\
		\label{A}
		&+ \left(\begin{array}{cc} \delta\alpha (1+\gamma_{eh}) & 2\gamma_{eh}\sqrt{\alpha_+\alpha_-} \\
			2\gamma_{eh}\sqrt{\alpha_+\alpha_-} & -\delta\alpha (1+\gamma_{eh}) \end{array}\right)\,,
\end{align}
such that
\begin{align}
		\label{U}
		U A U^\dagger = \left(\begin{array}{cc} \lambda_+ & 0 \\ 0 & \lambda_- \end{array}\right)\,,
\end{align}
with $\lambda_\pm=\alpha^2 + 2\gamma_{eh} \pm\lambda$.
	
The susceptibilities then follow straightforwardly from the diagonalized bosonic operators.
As we are only interested in the critical exponents within linear order in the coupling constant, we can express the charge bosonic fields as follows:
\begin{align}
		\label{varphi_expanded}
		\varphi^C_a &= \left(1+\frac{\gamma_{aa}}{2}\right)\tilde{\varphi}_a
		+ a \gamma_{eh}\frac{\alpha_a}{\delta\alpha}\tilde{\varphi}_{\bar{a}} \,,\\
		\label{vartheta_expanded}
		\vartheta^C_a &= \left(1-\frac{\gamma_{aa}}{2}\right)\tilde{\vartheta}_a
		+ a \gamma_{eh}\frac{\alpha_{\bar{a}}}{\delta\alpha}\tilde{\vartheta}_{\bar{a}} \,.
\end{align}
Furthermore, in leading order we can use $\tilde{v}_a\approx v_a$ since these corrections only influence prefactors but not the exponents.

The intra-pocket charge/spin susceptibility is then given by the following vacuum averages,
\begin{align}
  \nonumber
  \chi^{(1\rm D)}_{C/S,aa}(x) &= \frac{1}{(2\pi R_0)^2} \int d\tau \langle T e^{i\sqrt{2\pi}\varphi^C_a(\xi)}
                              e^{-i\sqrt{2\pi}\varphi^C_a(0)}\rangle \\
  \label{CS_intra_susceptibility}
                            &  \times \langle T e^{i\sqrt{2\pi}\varphi^S_a(\xi)}
                              e^{-i\sqrt{2\pi}\varphi^S_a(0)}\rangle \,.
\end{align}
The intra-pocket pair susceptibility is given by a similar equation with $\varphi^{C/S}_a\rightarrow\vartheta^{C/S}_a$. 
Using Eqs.~(\ref{varphi_expanded}), (\ref{vartheta_expanded}), we find the intra-pocket susceptibility scaling in leading order
\begin{align}
  \label{CS_intra}
  \chi^{(1\rm D)}_{C/S,aa}(x) &= \frac{1}{(2\pi R_0)^2}\int d\tau\left(\frac{R_0}{|x+iv_a\tau|}\right)^{2-\gamma_{aa}} \,,\\
  \label{P_intra}
  \chi^{(1\rm D)}_{P,aa}(x) &= \frac{1}{(2\pi R_0)^2}\int d\tau\left(\frac{R_0}{|x+iv_a\tau|}\right)^{2+\gamma_{aa}}\,.
\end{align}
After performing the integral over $\tau$, we then find the static susceptibilities,
\begin{align}
  \label{CS_static_intra}
  \chi^{(1\rm D)}_{C/S,aa}(x) &= \chi_{aa}^{(0)}(x) \left|\frac{x}{R_0}\right|^{\gamma_{aa}} \,,\\
  \label{P_static_intra}
  \chi^{(1\rm D)}_{P,aa}(x) &= \chi_{aa}^{(0)}(x)\left|\frac{x}{R_0}\right|^{-\gamma_{aa}}\,,
\end{align}
where $\chi_{aa}^{(0)}(x) = 1/(4\pi v_a |x|)$ is the noninteracting susceptibility and the cut-off $R_0$ is adjusted such that dressed and free susceptibilities are equal at $x = R_0$. This result is in perfect agreement with the RG approach.

The inter-pocket charge and spin susceptibilities are calculated in a similar fashion,
\begin{align}
  \nonumber
  \chi^{(1\rm D)}_{C/S,eh}(x) &= \frac{1}{(2\pi R_0)^2} \int d\tau\langle T e^{i\sqrt{\pi/2}\,\Phi^C_+(\xi)}
                              e^{-i\sqrt{\pi/2}\,\Phi^C_+(0)}\rangle \\
  \label{CS_inter_susceptibility}
                            & \times \langle T e^{i\sqrt{\pi/2}\,\Phi^S_+(\xi)}
                              e^{-i\sqrt{\pi/2}\,\Phi^S_+(0)}\rangle \,.
\end{align}
The inter-pocket pair susceptibility corresponds to $\Phi^{C/S}_+\rightarrow\Phi^{C/S}_-$, where
	\begin{align}
		\label{Phi_+}
		\Phi^{C/S}_+ &= \sum\limits_a \left(\varphi^{C/S}_a + a \vartheta^{C/S}_a\right) \,,\\
		\label{Phi_-}
		\Phi^{C/S}_- &= \sum\limits_a \left(a \varphi^{C/S}_a + \vartheta^{C/S}_a\right)\,.
	\end{align}
Using Eqs.~(\ref{varphi_expanded}), (\ref{vartheta_expanded}), we find
	\begin{align}
		\nonumber
		\Phi^{C}_+ &= \sum\limits_a \left(1+\frac{\gamma_{aa}}{2} - a \gamma_{eh}\frac{\alpha_{\bar{a}}}{\delta\alpha}\right)
		\tilde{\varphi}_a \\
		\label{Phi_+_expanded}
		& + \sum\limits_a a \left(1 - \frac{\gamma_{aa}}{2} + a \gamma_{eh} \frac{\alpha_a}{\delta\alpha}\right)
		\tilde{\vartheta}_a \,,\\
		\nonumber
		\Phi^{C}_- &= \sum\limits_a a \left(1+\frac{\gamma_{aa}}{2} + a \gamma_{eh}\frac{\alpha_{\bar{a}}}{\delta\alpha}\right)
		\tilde{\varphi}_a \\
		\label{Phi_-_expanded}
		& + \sum\limits_a \left(1 - \frac{\gamma_{aa}}{2} - a \gamma_{eh} \frac{\alpha_a}{\delta\alpha}\right)
		\tilde{\vartheta}_a \,.
	\end{align}
After straightforward calculations, we find the inter-pocket susceptibilities,
\begin{eqnarray}
  && \hspace{-25pt} \chi^{(1\rm D)}_{C/S,eh}(x) = \frac{1}{(2\pi R_0)^2} \nonumber \\
  && \hspace{20pt} \times \int d\tau\left[\frac{R_0^2}{(x+iv_e\tau)(x-iv_h\tau)}\right]^{1-\gamma_{eh}/2}
                              \,,  \label{CS_inter}\\
  && \hspace{-25pt} \chi^{(1\rm D)}_{P,eh}(x) = \frac{1}{(2\pi R_0)^2} \nonumber \\
  && \hspace{15pt} \times \int d\tau
                          \left[\frac{R_0^2}{(x+iv_e\tau)(x-iv_h\tau)}\right]^{1+\gamma_{eh}/2}
                          \,, \label{P_inter}
\end{eqnarray}
which, after integration over $\tau$, yield 
\begin{align}
  \label{CS_static_inter}
  \chi^{(1\rm D)}_{C/S,eh}(x) &= \chi_{eh}^{(0)}(x) \left|\frac{x}{R_0}\right|^{\gamma_{eh}} \,,\\
  \label{P_static_inter}
  \chi^{(1\rm D)}_{P,eh}(x) &= \chi_{eh}^{(0)}(x)\left|\frac{x}{R_0}\right|^{-\gamma_{eh}}\,,
\end{align}
with $\chi_{eh}^{(0)}(x)=1/(4\pi\bar{v} |x|)$. This agrees exactly with the  one-loop RG result  given in Eq.~(\ref{chiRG}).

Finally, we note that all results can be easily generalized to finite temperature $T > 0$. 
This corresponds to the following transformation in Eqs.~(\ref{CS_intra}), (\ref{P_intra}), (\ref{CS_inter}), (\ref{P_inter}):
\begin{align}
		\label{finite_T}
		x\pm iv_a\tau \rightarrow \frac{\beta v_a}{\pi} \sinh\left[\frac{\pi}{\beta v_a}\left(x\pm iv_a\tau\right)\right]\,,
\end{align}
where $\beta = 1/T$ is the inverse temperature, the Boltzmann constant $k_B = 1$.
We point out that all zero-temperature singularities in the relevant susceptibilities for $|\gamma_{ab}| > \gamma_c$ which we discussed in the main text are cut by any finite temperature.
This behavior is qualitatively different from the BCS-like mechanisms of instabilities which provide finite energy scales at $|\gamma_{ab}|>\gamma_c$ corresponding to finite critical temperatures.
In our situation, a forward-scattering interaction cannot stabilize any long-range order at any finite temperature, which is similar to 1D Luttinger liquids, where additional irrelevant interactions are necessary to stabilize long-range orders.

Below, we provide the temperature scaling of static susceptibilities.
All 1D static susceptibilities given by Eqs.~(\ref{CS_intra}), (\ref{P_intra}), (\ref{CS_inter}), (\ref{P_inter}) take the following form in case of a finite temperature $T > 0$, see the conformal transformation given by Eq.~(\ref{finite_T}),
\begin{eqnarray}
	&& \hspace{-20pt} \chi^{(1\rm D)}_{ab} (r) = \frac{T^2}{4 v_a v_b} \left(\frac{\sqrt{v_a v_b}}{\pi T R_0}\right)^{\gamma_0} \nonumber \\
	&& \hspace{-20pt} \times \! \int\limits_0^\beta \! \frac{d \tau}{\left[ \sinh\!\left(\frac{\pi T r}{v_a} + i \pi T \tau\right) \sinh\!\left(\frac{\pi T r}{v_b} - i \pi T \tau\right) \! \right]^{1 - \frac{\gamma_0}{2}}} , \label{staticT1}
\end{eqnarray}
where $\gamma_0 = \gamma_{ab}$ for charge and spin susceptibilities, $\gamma_0 = -\gamma_{ab}$ for pair susceptibility.
In order to take the integral over $\tau$, it is convenient to introduce a new complex variable $z = e^{2 i \pi T \tau} \rho$, such that the integration over $\tau$ is replaced by complex contour integral over the circle $S_\rho$ of radius $\rho = e^{\pi T r (v_b - v_a)/(v_a v_b)}$,
\begin{eqnarray}
	&& \hspace{-20pt} \chi^{(1\rm D)}_{ab} (r) = \frac{T}{v_a v_b} \left(\frac{\sqrt{v_a v_b}}{2 \pi T R_0}\right)^{\gamma_0} \nonumber \\
	&& \hspace{20pt} \times \oint\limits_{S_\rho} \frac{dz}{2 \pi i z} \left[\frac{z}{\left(z - \lambda\right) \left(\lambda^{-1} - z\right)}\right]^{1 - \frac{\gamma_0}{2}} , \label{staticT2}
\end{eqnarray}
where $\lambda = e^{-r/R_T} < \rho$.
Here, $R_T$ is a thermal length,
\begin{eqnarray}
	&& R_T = \frac{v_a v_b}{\pi (v_a + v_b) T} . \label{thermal}
\end{eqnarray}
The integral over $S_\rho$ in Eq.~(\ref{staticT2}) can be contracted to the integral over the branch cut located fully within the circle on the real axis $z \in (0, \lambda)$,
\begin{eqnarray}
	&& \hspace{0pt} \chi^{(1\rm D)}_{ab} (r) = \frac{T}{\pi v_a v_b} \left(\frac{\sqrt{v_a v_b}}{2 \pi T R_0}\right)^{\gamma_0} e^{-\frac{r}{R_T} \left(1-\frac{\gamma_0}{2}\right)} \sin \left(\frac{\pi \gamma_0}{2}\right) \nonumber \\ 
	&& \hspace{30pt} \times \int\limits_0^1 du \, u^{-\frac{\gamma_0}{2}} \left(1 - u \right)^{\frac{\gamma_0}{2} - 1} \left(1 - \lambda^2 u \right)^{\frac{\gamma_0}{2} - 1} ,
\end{eqnarray}
where we rescaled $z = \lambda u$, $u \in (0, 1)$.
Finally, the integral over $u$ corresponds to well known Euler representation of the hypergeometric function,
\begin{eqnarray}
	&& \hspace{-20pt} _2F_1(a,b;c;z) = \frac{\Gamma(c)}{\Gamma(b) \Gamma(c - b)} \nonumber \\
	&& \hspace{15pt} \times \int\limits_0^1 du \, u^{b - 1} (1 - u)^{c - b - 1} \left(1- z u\right)^{-a} .
\end{eqnarray}
Hence, the finite-temperature 1D susceptibilities take the following form:
\begin{eqnarray}
	&& \hspace{-15pt} \chi^{(1\rm D)}_{ab} (r) = \frac{e^{-\frac{r}{R_T} \left(1-\frac{\gamma_0}{2}\right)}}{\pi (v_a + v_b) R_T} \left(\frac{R_T}{\tilde{R}_0}\right)^{\gamma_0} \nonumber \\
	&& \hspace{25pt} \times \,  _2F_1\! \left(1 - \frac{\gamma_0}{2}, 1 - \frac{\gamma_0}{2}; 1; e^{- \frac{2 r}{R_T}} \right) . \label{staticTofr}
\end{eqnarray}
Here, we introduced an adjusted cut-off,
\begin{eqnarray}
	&& \tilde{R}_0 = R_0 \frac{2 \sqrt{v_a v_b}}{v_a + v_b} .
\end{eqnarray}
It is clear from Eq.~(\ref{staticTofr}) that the finite-temperature 1D static susceptibility is localized on the thermal scale $r \lesssim R_T$.
At zero-temperature, we found that the $D$-dimensional susceptibilities converged at $r \sim 1/\delta Q$, where $\delta Q$ is the distance from the corresponding anomaly, see Sec.~\ref{sec:Kohn}.
In case of finite temperature, we can set $\delta Q = 0$, and the corresponding convergence of $D$-dimensional Fourier transforms is provided on the scale $r \sim R_T$. 
Therefore, the temperature scaling of the $D$-dimensional susceptibilities can be obtained by the substitution,
\begin{eqnarray}
	&& \delta Q \to \frac{1}{R_T} \propto T . \label{QtoT}
\end{eqnarray}
Therefore, the finite-temperature static susceptibilities remain finite at any $T > 0$.
However, at $\gamma_0 > \gamma_c$ this temperature scaling is singular if $T \to 0$ which agrees with the non-Fermi-liquid behavior observed in many strongly correlated electron systems.

\section{Fourier transforms}
\label{app:fourier}

Here we provide the $D$-dimensional Fourier transforms of oscillatory functions with power-law amplitudes,
\begin{eqnarray}
&& \hspace{-15pt} \int d \bm r \, e^{-i \bm Q \cdot \bm r} \frac{\cos\left(Q_0 r - \phi\right)}{r^{D - \gamma}} \approx \left[\frac{2 \pi}{Q_0}\right]^{\gamma_c} \nonumber \\
&& \hspace{-10pt} \times \cos\left(\frac{\pi}{2} \left(\gamma - \gamma_c\right) + \left(\phi -\vartheta \right) \, \mathrm{sgn} \left(\delta Q\right)\right) \frac{\Gamma(\gamma - \gamma_c)}{|\delta Q|^{\gamma - \gamma_c}} , \label{Fourier}
\end{eqnarray}
where $\phi$ is an arbitrary phase, $\gamma_c = (D - 1)/2$ is given by Eq.~(\ref{gammac}),
$\vartheta = \pi (D - 1)/4$ is the semiclassical phase, $\delta Q = Q - Q_0 \ll Q_0$, $\textrm{sgn}(x)$ returns the sign of $x$.
Equation~(\ref{Fourier}) is the asymptotic expansion near $Q = Q_0$ that only takes into account the non-analytic $\propto |\delta Q|^{\gamma_c - \gamma}$ contribution, any analytic part can be safely discarded.
Next, we show a few steps of the derivation.

The angular integral over the $D$-dimensional solid angle $d \bm n$ can be expressed in terms of the Bessel function $J_\mu (z)$,
\begin{eqnarray}
&& \int d\bm n \, e^{-i \bm Q \cdot \bm r} = 2 \pi \left(\frac{2 \pi}{Q r}\right)^{\frac{D}{2} - 1} J_{\frac{D}{2} - 1} \left(Q r\right) . \label{angular}
\end{eqnarray}
Here we are interested in the asymptotic limit when $Q \approx Q_0$, so we consider the sector of large distances where $Q r \gg 1$,
\begin{eqnarray}
&& \int d\bm n \, e^{-i \bm Q \cdot \bm r} \approx  \left(\frac{2 \pi}{Q r}\right)^{\frac{D-1}{2}} 2 \cos \left(Q r - \vartheta \right) .
\end{eqnarray}
From the product $\cos \left(Q r - \vartheta \right) \cos \left(Q_0 r - \phi \right)$ we consider only the slowest $Q - Q_0$ harmonic that is sensitive to the long-distance limit, whereas the $Q + Q_0$ harmonic is discarded.
The integral over $r$ can be analytically continued to the Euler gamma function as follows,
\begin{eqnarray}
&& \int\limits_0^\infty \frac{dr}{r^{\gamma_c - \gamma + 1}} e^{i \delta Q r} = \frac{\Gamma(\gamma - \gamma_c)}{|\delta Q|^{\gamma - \gamma_c}} e^{\frac{i \pi}{2} (\gamma - \gamma_c) \mathrm{sgn} \left(\delta Q\right)} , \label{euler}
\end{eqnarray}
where the Wick rotation was performed, such that on the rotated contour we have $r = i z/\delta Q$, where $z \in (0, +\infty)$.
The formal short-distance singularity of the integral in Eq.~(\ref{euler}) at $r \to 0$ in case if $\gamma < \gamma_c$ is cured by the short-distance cut-off.
The ultraviolet divergence in this case just corresponds to the standard Fermi-liquid renormalizations coming from the short distances $r \sim \lambda_F$.
In contrast, the singularity for $\delta Q \to 0$ if $\gamma > \gamma_c$ indicates a strong infrared divergence of the integral at large distances that is only cut by finite $\delta Q$.

The Fourier transform of power-law functions without an oscillatory factor is the following,
\begin{eqnarray}
&& \hspace{-28pt} \int d \bm r \, \frac{e^{-i \bm Q \cdot \bm r}}{r^{D - \gamma}} = \left[\frac{2}{Q}\right]^\gamma \frac{\pi^{\frac{D}{2}}\Gamma\left(\frac{\gamma}{2}\right)}{\Gamma\left(\frac{D - \gamma}{2}\right)} . \label{powerlaw}
\end{eqnarray}
Equation~(\ref{powerlaw}) is exact at $\gamma > 0$.
It is straightforwardly derived using Eq.~(\ref{angular}) and the following integral representation of the Bessel function,
\begin{eqnarray}
&& \hspace{-7pt} J_\mu (u) = \frac{u^\mu}{2^\mu \sqrt{\pi} \Gamma\left(\mu + \frac{1}{2}\right)} \int\limits_{-1}^{1} ds \, \left(1 - s^2\right)^{\mu - \frac{1}{2}} e^{i s u} , \label{Bessel}
\end{eqnarray}
where $\mu > - 1/2$.
This integral representation is used to calculate the following integral,
\begin{eqnarray}
&& \hspace{-20pt}\int\limits_0^\infty du \, u^{\gamma - \frac{D}{2}} J_{\frac{D}{2} - 1} (u)  = \frac{\Gamma\left(\frac{\gamma}{2}\right)}{2^{\frac{D}{2} - \gamma} \Gamma\left(\frac{D - \gamma}{2}\right)} .
\end{eqnarray}

\end{document}